\documentclass[10pt]{article}
\usepackage{amssymb}
\usepackage{amsmath}
\usepackage{amsthm}
\usepackage{latexsym}
\usepackage[dvips]{epsfig}
\usepackage{mathrsfs}
\usepackage{xcolor}
\usepackage[bookmarks,colorlinks=false]{hyperref}
\usepackage{comment}

\theoremstyle{plain}
\newtheorem{proposition}{Proposition}
\newtheorem{remark}{Remark}
\newtheorem{lemma}{Lemma}
\newtheorem{theorem}{Theorem}

\newtheorem{definition}{Definition}

\newcounter{mnotecount}[section]

\newcommand{\mnotex}[1]%{}
{\protect{\stepcounter{mnotecount}}$^{\mbox{\footnotesize $\bullet$\themnotecount}}$ 
\marginpar{%\color{red}%
\raggedright\tiny\em
$\!\!\!\!\!\!\,\bullet$\themnotecount: #1} }

\begin{document}

\title{\textbf{Type D conformal initial data}}

\author{{\Large Alfonso Garc\'{\i}a-Parrado G\'omez-Lobo} 
	\thanks{E-mail address:
		{\tt alfonso@utf.mff.cuni.cz}
	} \\
	Institute of Theoretical Physics, Faculty of Mathematics and Physics,\\ 
	Charles University in Prague, V~Hole\v{s}ovi\v{c}k\'ach~2, 180~00 Praha 8, Czech 
	Republic
	\thanks{
		Current address: Departamento de Matem\'aticas,
		Campus de Rabanales, Universidad de C\'ordoba, 
		14071 C\'ordoba, Spain.
	} \\
}
\maketitle

\begin{abstract}
For a vacuum initial data set of the Einstein field equations it is possible to carry out 
a conformal rescaling or conformal compactification of the data giving 
rise to an initial data set for the Friedrich vacuum conformal equations. 
When will the data development with respect to the conformal equations of this set be a 
conformal extension of a type D solution? In this work we 
provide a set of necessary and sufficient 
conditions on a set of 
initial data for the conformal equations that guarantees that the data development of 
the conformal equations has a subset that is conformal to a vacuum type D solution of the Einstein's equations.
In particular we find the conditions under which this vacuum solution corresponds 
to the Kerr solution. Using our results we are able 
to show that there are no obstructions to extend the Petrov type of the physical spacetime to the 
unphysical spacetime if the conformal data are hyperboloidal.
\end{abstract}

\section{Introduction}

Since its introduction by Penrose \cite{ASYMPTOTICPRL,ZEROREST}, the notion of {\em conformal boundary} has found a wide number of applications in 
general relativity and theoretical physics. In general relativity the conformal boundary has been used to give a rigourous definition of 
{\em isolated system} (asymptotically simple space-time) 
and procedures to compute the total emission and absortion of gravitational radiation of such a system have been developed. 

The explicit computation of a conformal boundary with suitable properties for given exact solutions of the Einstein field equations is 
a difficult enterprise unless we deal with the simplest solutions. A possible approach is to set up the computation as {\em an initial value problem} 
for a system of hyperbolic equations involving a conformal rescaling of the metric tensor used in the Einstein's equations (the {\em physical metric}).
The main obstacle one needs to surmount by following this approach is the lack of conformal invariance of the Einstein's equations. This means that the
standard results that allow the formulation of the Einstein's equtions as a Cauchy problem do not apply after performing the conformal rescaling
and therefore one needs additional techniques to find hyperbolic equations for the conformally rescaled metric (the {\em unphysical metric}).

A hyperbolic formulation as described in the previous paragraph has been developed by Friedrich \cite{FRI81A,FRI81B} resulting in the so-called
{\em conformal field equations} and they have been sucessfully used to prove a 
number of remarkable global existence results: first proof \cite{FRI86A,FRI86B} of the {\em non-linear stability} 
of some of the simplest solutions of Einstein's equation 
(Minkowski and de Sitter) and similar results for the Einstein-Yang-Mills system \cite{FRI91} (see also \cite{LUBBE20121548}), 
purely radiative spacetimes \cite{JUANLUBBERADIATIVE},
cosmological solutions \cite{LUBBE20131} and the asymptotic region of the Schwarzschild-de Sitter black hole \cite{GASPERIN2017}.
In any case, the rough idea is that the hyperbolic character of the conformal equations makes 
it possible to use classical local existence results of 
the partial differential equations theory to prove a local existence result for the former. 
The conformal relation between the unphysical
metric and the physical one, and the knowledge of the geometric properties of the conformal 
boundary enables us to turn a local existence 
result in the unphysical space-time into a global existence result for the physical space-time. 
Other very important global existence results where the conformal equations have played a key role
deal with {\em asymptotically simple} spacetimes with a smooth conformal boundary. See 
\cite{CUTLER89,CHRUSCIELDELAY02} for examples of this situation.

Given a set of hyperbolic equations or a {\em hyperbolic reduction} of a set of tensorial equations 
it makes sense to investigate {\em its initial data problem}. This is a set of conditions on an initial data hypersurface 
or Cauchy hypersurface ensuring the existence of a solution of the hyperbolic system.
Formulations of the initial value problem for the conformal equations can be found in the above references and also in \cite{FRI86A}.
In this framework different kinds of initial data have been studied: the asymptotic characteristic initial value problem  
\cite{FRI81A,FRI81B}, data prescribed at (spacelike) past null infinity \cite{FRI86B}, 
data for space-times with a timelike conformal boundary \cite{FRIADS95} 
and data for Kerr-de Sitter spacetimes at null infinity \cite{MSENOTORB16}. See \cite{JUANBOOK,FRAUENDIENERREVIEW} 
for a detailed review and information about all these topics.

Suppose that we have a vacuum initial data set for the Einstein field equations 
and carry out a conformal rescaling (conformal compactification) 
of the data. 
This gives rise to initial data for the (vacuum) conformal equations.
When will the data development with respect to the conformal equations be a conformal extension of a 
vacuum type D solution? This work provides an answer to this question that is written exclusively
in terms of the quantities used to define an initial data set of the conformal equations.
We also show that for hyperboloidal data  
there are no obstructions to the extension of the Petrov type of the physical space-time
to the unphysical space-time (see Theorem \ref{theo:data-at-boundary} for more details).

The {\em type D conformal initial data} are
a set of conditions that should be appended to the {\em conformal constraint equations} and 
therefore given exclusively in terms of the data of the conformal equations. We also prove that the set of conditions 
is a set of {\em necessary and sufficient} conditions, so any other initial data set for the conformal equations 
whose development admits a subset that is conformal to a Petrov type D vacuum solution must be already dependent from our set
in some region of the initial data hypersuface.
The method presented in this paper to construct initial data for the conformal equations is valid 
for any vacuum Petrov type D solution and we 
also particularize it for the case of the Kerr solution. 
Note that the construction of initial data for the conformal equations corresponding to Kerr data 
is the starting point in order to study the non-linear stability of the 
Kerr black hole using conformal techniques, in the spirit of the results described above. 
Note also that once an initial value problem for the conformal
field equations has been set up, a local existence result of the conformal hyperbolic system 
may translate into a global existence result for the 
original Einstein's equations, provided some extra conditions are met. 
In general a local existence result is far easier to obtain than a global one, so the use of conformal techniques 
could play an important role in the solution of the non-linear stability of the Kerr black hole.
Also the initial data so constructed could be used as the starting point in the analysis of the conformal boundary 
properties for members of this 
important class of exact solutions. In this sense there are already results for the Schwarzschild \cite{FRIEDRICHCF2003} and 
the Kottler family of solutions \cite{CONFORMALGEODESICSKOTTLER} where 
the construction of congruences of {\em conformal geodesics} enables us to determine geometric properties of the conformal boundary 
without carrying out the actual conformal extension. 

This paper is structured as follows: in section \ref{sec:conformal-eqs} we recall the formulation of the vacuum conformal equations 
and the construction of conformal initial data sets for them. 
Section \ref{sec:type-D} reviews an invariant characterization of Petrov type D solutions 
needed for the construction of initial data for the conformal equations. This is the subject of section \ref{sec:conformal-data-D} 
where the main results of this paper (Theorems \ref{theo:conformal-petrov-d} and \ref{theo:conformal-petrov-d-necessary}) are presented. 
In section \ref{sec:kerr-conformal-data} we particularize these results to the case in which the data are constructed from data 
for the Kerr solution (Theorems \ref{theo:kerr-conformal-data} and \ref{theo:kerr-conformal-data-necessary}).
Section \ref{sec:conformal-limit} analyzes the conformal boundary limit of the initial data conditions obtained in the previous sections  
finding that there are no obstructions to the extension of the data through the conformal boundary whenever
the data are {\em hyperboloidal}.
We discuss possible applications in section \ref{sec:conclusions}.

All the tensor computations in this paper have been carried out with the system {\em xAct} \cite{XACT}, 
a {\em Wolfram Language} suite for doing tensor analysis (see also  
\cite{XPERM}).

\section{The vacuum Friedrich conformal equations and their initial data}
\label{sec:conformal-eqs}
Let $(\tilde{\mathcal{M}},{\tilde g}_{ab})$ be a 4-dimensional Lorentzian manifold (physical space-time) and
$(\mathcal{M},g_{ab})$ a se\-cond Lorentzian manifold (unphysical spacetime) which is {\em conformally related} to the 
first in the following fashion (the signature convention for both metrics is $(-,+,+,+)$) 
\begin{equation}
g_{ab} = \Theta^2 \tilde{g}_{ab}.
\label{eq:unphysicaltophysical-downstairs}
\end{equation}
In the previous relation a conformal map (conformal embedding) 
from $\tilde{\mathcal{M}}$ to $\mathcal{M}$ is understood and the 
conformal factor $\Theta$ is assumed to be a smooth function which does not vanish 
in the manifold $\tilde{\mathcal{M}}$.
We use small Latin letters to denote abstract indices 
of tensors in $\mathcal{M}$ and $\tilde{\mathcal{M}}$. 
Indices are always raised and lowered with respect to the unphysical metric $g_{ab}$ 
with the exception of $\tilde{g}^{ab}$ where we follow 
the traditional convention that it represents the inverse 
of $\tilde{g}_{ab}$
\begin{equation}
\tilde{g}^{ac}\tilde{g}_{cb}=\delta_b{}^a.
\end{equation}
Hence, the explicit relation between 
the physical and the unphysical contravariant metric tensors is then
\begin{equation}
g^{ab} = \frac{\tilde{g}^{ab}}{\Theta^2}.
\label{eq:unphysicaltophysical-upstairs}
\end{equation}

Each of the metric tensors $g_{ab}$, $\tilde{g}_{ab}$ has its own volume element,
denoted respectively by $\eta_{abcd}$ and $\tilde{\eta}_{abcd}$.
Using (\ref{eq:unphysicaltophysical-downstairs}) we deduce the relation
\begin{equation}
\tilde{\eta}_{abcd} = \frac{\eta_{abcd}}{\Theta^4}.
\label{eq:etaunphystoetaphys}
\end{equation}
Also each metric tensor has its own Levi-Civita connection denoted respectively 
by $\nabla_a$, $\tilde{\nabla}_a$ which are used to define the connection coefficients and the curvature 
tensors in the standard fashion. Our conventions for the (unphysical) Riemann, Ricci and Weyl tensors 
are
\begin{equation}
\nabla_{a}\nabla_{b}\omega_{c} -  \nabla_{b}\nabla_{a}\omega_{c} =  R_{abc}{}^{d} \omega_{d}\;,
\label{eq:define-riemann}
\end{equation}
\begin{equation}
 R_{ac}\equiv R_{abc}{}^b\;,
\label{eq:define-ricci}
\end{equation}
\begin{equation}
C_{abcd}\equiv R_{abcd} - 2 L_{d[b}g_{a]c} - 2 g_{d[b}L_{a]c}\;,
\label{eq:define-weyl}
\end{equation}
where the unphysical Schouten tensor is defined by
\begin{equation}
 L_{ab} \equiv \tfrac{1}{2} (R_{ab} -  \tfrac{1}{6} R g_{ab}).
\label{eq:define-schouten}
\end{equation}
The conventions for the corresponding physical quantities are similar and we use a tilde 
over the symbol employed for a unphysical spacetime tensor to denote its physical counterpart. The only exception
of this rule occurs for the physical Weyl tensor, where the notation is $\tilde{W}_{abcd}$ 
(see eq. (\ref{eq:weylunphystoweylphys}) below).
Recall that the Riemann, Ricci and Weyl tensors have a {\em natural index configuration} in their definition 
which is important to bear in mind when working with two different metric tensors. This is so because a 
tensorial expression containing any of these tensors in a non-natural index configuration requires 
a clear convention telling us the metric which was used to change from the natural index configuration 
to the non-natural one. In this sense, eqs. (\ref{eq:define-riemann})-(\ref{eq:define-schouten})
present the Riemann, Ricci, Weyl and Schouten tensors in their natural index configuration.
As already mentioned we are adopting the convention of taking the unphysical metric as the 
metric used to raise and lower indices and therefore, this shall be the metric we are going to use
to change the natural index configuration of any tensor.

Standard computations enable us to find the relations between the connection coefficients and curvature
tensors of $\nabla$ and $\tilde{\nabla}$. For us the relation between the unphysical 
Weyl tensor $C_{abcd}$ and the physical one $\tilde{W}_{abcd}$ will be specially important 
\begin{equation}
C_{abcd} = \Theta^2 \tilde{W}_{abcd}.
\label{eq:weylunphystoweylphys}
\end{equation}

The star $*$ is used to denote both the Hodge dual and the complex conjugation and we leave to the context 
the distinction between these two.

\subsection{The metric conformal equations}
An interesting situation occurs when the unphysical space-time is conformally related to a physical 
space-time which is a vacuum solution of the Einstein equations. In this case it is a non-trivial problem 
to find a set of hyperbolic field equations involving the unphysical metric and {\em regular} when the conformal 
factor $\Theta$ vanishes. Under suitable gauge choices and conditions,
the set of {\em metric conformal field equations} yields a hyperbolic system with these properties.
In this sense we can say that the metric conformal field equations are 
a {\em regular conformal representation} of the Einstein field equations.
Suppose that the physical space-time fulfills the vacuum Einstein equations with cosmological constant
$\lambda$
\begin{equation}
 \tilde{R}_{ab} = \lambda \tilde{g}_{ab}\;,
\label{eq:vacuum-lambda}
\end{equation}
then the {\em vacuum metric conformal equations} hold in the unphysical space-time $(\mathcal{M},g_{ab})$ 
\begin{subequations}
\begin{eqnarray}
&&\Sigma_{a} = \nabla_{a}\Theta\;,\label{eq:cfe0}\\
&&\nabla_{b}\nabla_{a}\Theta = - \Theta L_{ab} + s g_{ab}\;,\label{eq:cfe1}\\
&&\nabla_{a}s = - L_{ab} \nabla^{b}\Theta\;,\label{eq:cfe2}\\
&&\nabla_{a}L_{bc} -  \nabla_{b}L_{ac} = - d_{abcd} \nabla^{d}\Theta\;,\label{eq:cfe3}\\
&&\nabla_{p}d_{abc}{}^{p} = 0\;,\label{eq:cfe4}
\end{eqnarray}
\end{subequations}
together with the {\em constraint}
\begin{equation}
\lambda = 6 \Theta s - 3 \nabla_{a}\Theta \nabla^{a}\Theta.
\label{eq:cfe-constraint}
\end{equation}
In the formulation of the conformal field equations we have introduced the
{\em Friedrich scalar}
\begin{equation}
s \equiv \tfrac{1}{24} \Theta R + \tfrac{1}{4} g^{ab} \nabla_{b}\nabla_{a}\Theta\;,
\end{equation}
and the {\em rescaled Weyl tensor}
\begin{equation}
d_{abcd} \equiv \frac{C_{abcd}}{\Theta}.
\label{eq:rescaledweyl}
\end{equation}
The vacuum conformal equations give rise to a hyperbolic system in the unphysical manifold 
$\mathcal{M}$ for the following variables (see \cite{FRI86A,FRI86B})
\begin{equation}
 \Theta\;,\quad \Sigma_{a}\;,\quad s\;,\quad L_{bc}\;,\quad d_{abcd}\;,\quad g_{ab}.
\end{equation}

We recall for later use the following result (see Theorem 3.1 of \cite{FRIEDRICH83})
\begin{proposition}
 If 
 \begin{equation}
 \Theta\;,\quad \Sigma_{a}\;,\quad s\;,\quad L_{bc}\;,\quad d_{abcd}\;,\quad g_{ab}.
\end{equation}
is a solution of (\ref{eq:cfe0})-(\ref{eq:cfe4}) such that $\Theta\neq 0$ on an open set $\mathcal{U}\subset\mathcal{M}$
and (\ref{eq:cfe-constraint}) is fulfilled at least a point $p\in\mathcal{M}$ then the metric $\tilde{g}_{ab}$ is a solution 
of (\ref{eq:vacuum-lambda}).
\label{prop:conformal-solution}
\end{proposition}

\subsection{The initial data problem for the metric conformal equations}
To prescribe initial data for the conformal equations we follow the standard approach of 
defining a spacelike Cauchy hypersurface $\mathcal{S}\subset\mathcal{M}$. $\mathcal{S}$ is an embedded Riemannian manifold 
endowed with a Riemannian metric (we shall use the same symbol $\mathcal{S}$ for the Riemannian manifold and 
its image in $\mathcal{M}$ under the embedding if no confusion arises).
We define next a foliation $\{\mathcal{S}_t\}$, $t\in I\subset\mathbb{R}$ of the unphysical space-time $\mathcal{M}$
such that the leaves $\mathcal{S}_t\subset\mathcal{M}$ are spacelike 
hypersurfaces. Furthermore the foliation is chosen in such a way that $\mathcal{S}_0=\mathcal{S}$. The foliation can be characterized by any unit integrable timelike 
vector field $n^a$ defined on $\mathcal{M}$ which is orthogonal to the leaves.  
We can use $n^a$ to introduce the {\em spatial metric}
\begin{equation}
 h_{ab}\equiv g_{ab}+n_an_b.\
\end{equation}
The spatial metric enables us to define {\em spatial tensors} on $\mathcal{M}$ in the standard way. 
The embedding of $\mathcal{S}$ into $\mathcal{M}$ sets a 
one-to-one correspondence between spatial tensors on $\mathcal{M}$ and tensor fields on $\mathcal{S}$ 
and for that reason we shall use the same set of abstract indices for tensorial quantities on $\mathcal{S}$ as for those 
in $\mathcal{M}$.
Indices of tensorial quantities on $\mathcal{S}$ are always raised and lowered with the metric $h_{ab}$. A very important spatial tensor
is the extrinsic curvature defined by (\ref{eq:cf-data2}) below
$$
K_{ab}\equiv h_{a}{}^{c} h_{b}{}^{d} \nabla_{c}n_{d}.
$$
Given the existence of a conformal map between $\mathcal{M}$ and $\tilde{\mathcal{M}}$ one can relate $\mathcal{S}$, $\{\mathcal{S}_t\}$, $n^a$
and $h_{ab}$ to quantities defined in the same fashion on $\tilde{\mathcal{M}}$ using the metric $\tilde{g}_{ab}$.  In this way 
we introduce the vector field $\tilde{n}^a$ (physical normal) which is normalized with respect to $\tilde{g}_{ab}$, the Riemannian manifold
$\tilde{S}$, the spatial metric $\tilde{h}_{ab}$ and the extrinsic curvature $\tilde{K}_{ab}$. 
Since $\tilde{\mathcal{M}}$ can be conformally embedded in $\mathcal{M}$ we may write $\tilde{\mathcal{S}}\subset\mathcal{S}$. 
The relation between $n^a$ and $\tilde{n}^a$ is (recall that indices are raised and lowered with the unphysical metric and that the push-forward and 
pull-back by a conformal map between $\mathcal{M}$ and $\tilde{\mathcal{M}}$ is understood)
\begin{equation}
 n^a=\frac{\tilde{n}^a}{\Theta}\;,\quad n_a=\Theta \tilde{N}_a\;,\quad \tilde{N}_a\equiv\tilde{g}_{ab}\tilde{n}^b.
\label{eq:normal-unphys-to-normal-phys}
\end{equation}
It can be easily seen that the conditions (\ref{eq:normal-unphys-to-normal-phys}) 
entail the following relations 
\begin{equation}
h_{ab} = \Omega^2 \tilde{h}_{ab}\;,\quad
K_{ab} = \Omega (\tilde{K}_{ab} + \sigma\tilde{h}_{ab}).
\label{eq:hKToTildehK}
\end{equation}
These relations can be inverted yielding
\begin{equation}
\tilde{h}_{ab} = \frac{h_{ab}}{\Omega^2}\;,\quad
\tilde{K}_{ab} = \frac{K_{ab}}{\Omega} -  h_{ab}\frac{\sigma}{\Omega^2}.
\label{eq:TildehKTohK}
\end{equation}
The indices of spatial tensors can be raised and lowered with a spatial metric. 
Consistent with our convention for raising and lowering of space-time index tensors, we shall
use the spatial metric $h_{ab}$ for index raising and lowering.
Again, the only exception to this convention is $\tilde{h}^{ab}$ that
is the inverse of $\tilde{h}_{ab}$ and thus we have
\begin{equation}
\tilde{h}^{ab}\equiv\Omega^2 h^{ab}\;,\quad
\tilde{h}^{ac}\tilde{h}_{cb}=h^{ac}h_{cb}=h^a{}_c\;,\quad
\end{equation}

We can take any of the previous vector fields as the starting point to carry out a standard 1+3 decomposition 
(see \cite{SE-DYNAMICALAWS} and references therein). 
Since we are working with the conformal field equations which are formulated 
in terms of the unphysical metric, we choose to carry out the 1+3 decomposition
using the unphysical normal $n^a$ and the unphysical spatial metric $h_{ab}$. 
In this way, and following \cite{DIEGOJUANADS2018,JUANDIEGO2018}, we introduce the so-called 
{\em initial data quantities} for the conformal equations defined as follows
\begin{eqnarray}
&& h_{ab}\equiv n_{a} n_{b} + g_{ab}\;,\mbox{(first fundamental form of the embedded manifold $\mathcal{S}$).}\;,\label{eq:cf-data1}\\
&& K_{ab}\equiv h_{a}{}^{c} h_{b}{}^{d} \nabla_{c}n_{d}\;,\mbox{(second fundamental form of the embedded manifold $\mathcal{S}$).}\label{eq:cf-data2}\\
&& \Omega\;,\mbox{scalar function on $\mathcal{S}$ (restriction of $\Theta$ to the embedded manifold $\mathcal{S}$).}\label{eq:cf-data3}\\
&& \sigma\;,\sigma_a\;,\mbox{Scalar and vector field on $\mathcal{S}$ defined from the orthogonal splitting of $\Sigma_a$:}
\nonumber\\ 
&&\Sigma_{a} = - n_{a} \sigma + \sigma_{a}\Longrightarrow \sigma_a=D_a\Omega.\label{eq:cf-data5}\\
&&l_{ab} = r_{ab} -  \tfrac{r}{4} h_{ab}\;, 
\mbox{ Schouten tensor with respect to the Riemannian metric $h_{ab}$.}\label{eq:cf-data6}
\end{eqnarray}
Here we have defined the Levi-Civita connection $D_a$ compatible with the spatial metric $h_{ab}$ in the standard way.
From these fundamental quantities we construct the following {\em derived initial data quantities} on $\mathcal{S}$
(we indicate in brackets their correspondence with the space-time tensors)
\begin{subequations}
\begin{eqnarray}
&&s \equiv \frac{1}{3} \biggl(\frac{\Omega}{4} (K^2 + r - K_{ac} K^{ac}) -  K^{b}{}_{b}\sigma + D_{b}\sigma^{b}\biggr) \;,
\mbox{(restriction of the Friedrich scalar to $\mathcal{S}$\;),}\nonumber\\
&&\label{eq:fund-quantity-1}\\
&&\theta_{ab} \equiv \frac{1}{\Omega}(s h_{ab} + \sigma K_{ab} - D_{b}\sigma_{a})\;,\;\;
\mbox{(spatial part of $L_{ab}$)}\;,
\label{eq:fund-quantity-2}\\
&& \theta_{b}\equiv \frac{1}{\Omega}(K^{a}{}_{b} \sigma_{a} -  D_{b}\sigma)\;,
\mbox{(transversal part of $L_{ab}$)}\;,\;\;
\label{eq:fund-quantity-3}\\
&& d_{ac} \equiv \frac{1}{4 \Omega}\left((K_{db} K^{db} -  K^2) h_{ac}+4(K_{ac} K - K_{a}{}^{d} K_{cd} + l_{ac} 
- \theta_{ac})\right)\;,\nonumber\\
&&\mbox{(electric part of $d_{abcd}$)}\;,
\label{eq:fund-quantity-4}\\
&& d_{bdc}\equiv \frac{2}{\Omega}(h_{b[c} \theta_{d]} +  D_{[d}K_{c]b})\;,\label{eq:fund-quantity-5}\\
&& d^{*}{}_{ae} = \tfrac{1}{2} \eta_{ecdp} d_{a}{}^{dp} n^{c}\;,\;\;
\mbox{(magnetic part of $d_{abcd}$)}.
\label{eq:fund-quantity-6}
\end{eqnarray}
\end{subequations}
The initial data quantities fulfill the 
{\em conformal constraint equations} \cite{DIEGOJUANADS2018,JUANDIEGO2018,JUANBOOK}
\begin{subequations}
\begin{eqnarray}
&&\lambda = 6 \Omega s + 3 \sigma^2 - 3 \sigma_{a} \sigma^{a}\;,\label{eq:conf-constrant-lambda}\\
&& D_{a}s =  \theta_{a} \sigma -  \theta_{ab} \sigma^{b}\;,\label{eq:conf-constrant-3}\\
&& D_{a}\theta_{bc} -  D_{b}\theta_{ac} = -2\theta_{[a}K_{b]c}+
d_{cab} \sigma +  2\sigma_{[a} d_{b]c} - 2 d_{d[b} h_{a]c} \sigma^{d}\;,\label{eq:conf-constrant-4}\\
&&D_{a}\theta_{b} -  D_{b}\theta_{a} = d_{cab} \sigma^{c}\;,\label{eq:conf-constrant-5}\\
&&D_{a}d^{*}{}_{k}{}^{a} = d^{bc} \eta_{kacj} K_{b}{}^{j} n^{a}\;,\label{eq:conf-constrant-6}\\
&&D_{c}d_{b}{}^{c} = -\eta_{bcde} K^{ac} d^{*}{}_{a}{}^{e} n^{d}\;,\label{eq:conf-constrant-7}
\end{eqnarray}
\end{subequations}
Eq. (\ref{eq:conf-constrant-lambda}) is the spatial part of (\ref{eq:cfe-constraint}), 
(\ref{eq:conf-constrant-3}) is the spatial part of (\ref{eq:cfe2}), 
(\ref{eq:conf-constrant-4})-(\ref{eq:conf-constrant-5}) are the spatial part of 
(\ref{eq:cfe3}) and (\ref{eq:conf-constrant-6})-(\ref{eq:conf-constrant-7})
are the spatial part of (\ref{eq:cfe4}).

Possible initial data for the conformal equations are given by the  
{\em conformal hyperboloidal initial data sets} as introduced in \cite{FRIEDRICH83}.
\begin{definition}[vacuum conformal hyperboloidal initial data]
Let $(\tilde{\mathcal{S}},\tilde{h}_{ab},\tilde{K}_{ab})$ a vacuum initial data set (see \ref{theorem:vacuum-data})
and define from it the set $\mathcal{S}$ and the quantities 
\begin{equation}
{\mathcal C}\equiv(\mathcal{S},h_{ab},K_{ab},\Omega,\sigma,\sigma_a,s,\theta_{ab},\theta_b,d_{ab},d_{abc})\;,
\label{eq:conformal-hyperboloidal-data}
\end{equation}
in the manner explained in the previous paragraphs.
A vacumm conformal hyperboloidal initial data set is an initial data set in which $\mathcal{S}$ is a manifold diffeomorphic 
to the closed unit ball in $\mathbb{R}^3$ whose boundary
is denoted by $Z$. One has then that $\tilde{\mathcal{S}}$ is defined by $\tilde{\mathcal{S}}\equiv S\setminus Z$ and also
the following additional requirements
\begin{enumerate}
 \item $\Omega>0$ on $\tilde{\mathcal{S}}$.
 \item $\Omega=0$ on $Z$ and if $\lambda=0$ then $\sigma^2-\sigma_a\sigma^a=0$, $\sigma>0$ on $Z$.
 \item The set of quantities ${\mathcal C}$ fulfills the conformal constraint 
 equations (\ref{eq:conf-constrant-lambda})-(\ref{eq:conf-constrant-7}).
% \item On $\tilde{\mathcal{S}}$, one has the relations \mnotex{Does this appear in the original definition?}
% \begin{equation}
% h_{ab} \equiv \Omega^2 \tilde{h}_{ab}\;,\quad
% K_{ab} \equiv \Omega(\tilde{K}_{ab} + \sigma\tilde{h}_{ab}).
% \end{equation}
\end{enumerate}
\label{def:conformal-initial-data}
\end{definition}
We recall now the following result for a vacuum conformal hyperboloidal initial data set 
(see \cite{FRIEDRICH83} for a proof). 
\begin{theorem}[Hyperboloidal initial data for the vacuum conformal equations]
For a smooth vacuum conformal hyperboloidal initial data set defined by (\ref{eq:conformal-hyperboloidal-data})  
there exists a solution of the conformal equations (\ref{eq:cfe0})-(\ref{eq:cfe4}) 
\begin{equation}
 \Theta\;,\quad \Sigma_{a}\;,\quad s\;,\quad L_{bc}\;,\quad d_{abcd}\;,\quad g_{ab}\;,
\end{equation}
such that $g_{ab}=\Theta^2 \tilde{g}_{ab}$ where $\tilde{g}_{ab}$ is the vacuum solution of
the Einstein's equations corresponding to the initial data
$(\tilde{\mathcal{S}},\tilde{h}_{ab},\tilde{K}_{ab})$ used to construct the conformal hyperboloidal initial data set.
\label{theo:conf-constraints}
\end{theorem}

In our case, we can simplify the analysis of the conformal constraint equations 
(\ref{eq:conf-constrant-lambda})-(\ref{eq:conf-constrant-7}) by means of the following result
(see Lemma 11.1 of \cite{JUANBOOK})
\begin{theorem}
The set $\mathcal{C}$ fulfills the conformal constraint equations 
(\ref{eq:conf-constrant-lambda})-(\ref{eq:conf-constrant-7}) if and 
only if \\
$\{h_{ab},K_{ab},\Omega,\sigma,\sigma_a\}$ fulfills the conformal Hamiltonian and momentum constraints on $\mathcal{S}$:
\begin{subequations}
\begin{eqnarray}
&&2 \lambda = \Omega^2 (r + K^2 - (K_{ab} K^{ab})) - 6 (\sigma_{a} \sigma^{a}) + 4 \Omega (D_{a}\sigma^{a}) 
- 4\Omega K \sigma + 6 \sigma^2\;,\label{eq:conf-hamiltonian-constraint}\\
&& \Omega(D_{b}K_{d}{}^{b} - D_{d}K^{b}{}_{b}) = 2( \sigma_{a} K^{a}{}_{d} -  D_{d}\sigma).\label{eq:conf-momentum-constrant}
\end{eqnarray}
\end{subequations}
\label{theo:conf-ham-mom}
\end{theorem}

\section{An invariant characterization of the Petrov type D condition}
\label{sec:type-D}
From the physical metric $\tilde{g}_{ab}$ and its inverse $\tilde{g}^{ab}$, we define the volume element $\tilde{\eta}_{abcd}$, 
the Weyl tensor $\tilde{W}_{abcd}$, its right dual $\tilde{W}^*_{abcd}$ and the self-dual Weyl tensor
\begin{equation}
\tilde{\mathcal{W}}_{abcd}\equiv\frac{1}{2} (\tilde{W}_{abcd} - i \tilde{W}^{*}{}_{abcd}).
\end{equation}
We define the physical Weyl scalars
\begin{eqnarray}
&&\tilde{\mathit{a}} \equiv \tilde{g}^{ac}\tilde{g}^{be}\tilde{g}^{fq}\tilde{g}^{pd} \tilde{\mathcal{W}}_{abpf} \tilde{\mathcal{W}}_{cedq}\;,\\
&&\tilde{\mathit{b}} \equiv \tilde{g}^{ai}\tilde{g}^{bj}\tilde{g}^{ce}\tilde{g}^{dg}\tilde{g}^{pf}\tilde{g}^{qh}\tilde{\mathcal{W}}_{abcd} \tilde{\mathcal{W}}_{egpq} 
\tilde{\mathcal{W}}_{fhij}\;,\\
&&\tilde{\mathit{w}}\equiv-\frac{\tilde{\mathit{b}}}{2\tilde{\mathit{a}}}\;,
\end{eqnarray}
and the tensors
\begin{equation}
\tilde{G}_{abmh} \equiv \tilde{g}_{am} \tilde{g}_{bh} -  \tilde{g}_{ah} \tilde{g}_{bm}\;,\quad
\tilde{\mathcal{G}}_{abcd} \equiv \tfrac{1}{2} (-i \tilde{\eta}_{abcd} + \tilde{G}_{abcd}).
\end{equation}

We recall the following results from \cite{FERRANDOCOVARIANTPETROVTYPE,FERRSAEZTYPED,AGPTYPEDDATA}
\begin{theorem}
The physical spacetime $(\tilde{\mathcal{M}},\tilde{g}_{ab})$ is of ``genuine'' Petrov type D (Petrov type D, but not any of its specializations) if and only if 
\begin{equation}
\tilde{a}\neq 0\;,\quad \tilde{\mathcal{D}}_{abcd}=0\;,
\end{equation}
where
\begin{equation}
\tilde{\mathcal{D}}_{abhe}\equiv \tilde{\mathcal{W}}_{abcd}\tilde{g}^{cp} 
\tilde{g}^{dq}\tilde{\mathcal{W}}_{pqhe} -  
\frac{\tilde{\mathit{a}}}{6}\tilde{\mathcal{G}}_{abhe} - 
\frac{\tilde{\mathit{b}}}{\tilde{\mathit{a}}}\tilde{\mathcal{W}}_{abhe}.
\end{equation}
\label{theo:type-D}
\end{theorem}
In this work we shall only be concerned with those Petrov type D solutions characterized by Theorem \ref{theo:type-D}.  
\begin{proposition}
 Under the conditions of Theorem \ref{theo:type-D} and assuming that (\ref{eq:vacuum-lambda}) holds, we have that the 1-form $\tilde{\xi}_a$ defined by the equation
\begin{equation}
 \tilde{\Xi}_{ac} = \tilde{\varphi}\tilde{\xi}_{a} \tilde{\xi}_{c}\;,
\label{eq:invariant-killing}
\end{equation}
fulfills the condition
\begin{equation}
 \tilde{\nabla}_a\tilde{\xi}_b + \tilde{\nabla}_b\tilde{\xi}_a = 0\;,
\end{equation}
where
\begin{eqnarray}
	&&\tilde{\varphi} \equiv \frac{27}{2} \tilde{\mathit{w}}^{11/3}\;,\\
	&&\tilde{\Xi}_{ac} \equiv (
		\tilde{g}^{bp} \tilde{g}^{dq} \tilde{\mathcal{W}}_{apcq} -  
		\tilde{\mathit{w}}\tilde{\mathcal{G}}_{apcq} 
		\tilde{g}^{bp} 
		\tilde{g}^{dq}
	) 
	\tilde{\nabla}_{b}\tilde{\mathit{w}}\tilde{\nabla}_{d}\tilde{\mathit{w}}
\end{eqnarray}
\label{prop:killing-candidate}
\end{proposition}

All the previous results have a counterpart formulated with respect to the unphysical metric $g_{ab}$. 
To find the corresponding formulations we need to make similar definitions for the symbols used in 
Theorem \ref{theo:type-D} and Proposition \ref{prop:killing-candidate}, but now using the unphysical metric instead of the 
physical one. The notation for the new symbols so defined is obtained by just removing the tildes over the symbols 
used in the physical space-time.

\subsection{Conformal rescaling of the Petrov type D conditions in vacuum}
\begin{proposition}
The relation between the physical quantities defined in Theorem \ref{theo:type-D} 
and Proposition \ref{prop:killing-candidate} and the corresponding unphysical ones is given by
\begin{subequations}
\begin{eqnarray}
&&\tilde{\mathit{a}} = \Theta^6 \mathit{a}\;,\\
&&\tilde{\mathit{b}} = \Theta^9 \mathit{b}\;,\\
&&\tilde{\mathit{w}} = \Theta^3 \mathit{w}\;,\\
&&\widetilde{G}_{abcd} = \frac{G_{abcd}}{\Theta^4}\;,\\
&&\tilde{\mathcal{G}}_{abcd} = \frac{\mathcal{G}_{abcd}}{\Theta^4}\;,\\
&&\mathfrak{d}_{abcd} = \Theta \tilde{\mathcal{W}}_{abcd}\;,\\
&&\widetilde{\mathcal{D}}_{abcd}=\Theta^2\mathcal{D}_{abcd}\;,\label{eq:conformal-rescaling-D}\\
&&\tilde{\varphi} = \Theta^{11} \varphi\;,\label{eq:conformal-rescaling-varphi}\\
&& \tilde{\Xi}_{ac} = \Theta^3 (\mathfrak{d}_{a}{}^{b}{}_{c}{}^{d}-\mathcal{G}_{a}{}^{b}{}_{c}{}^{d} \mathit{w})
\nabla_{b}\tilde{\mathit{w}} \nabla_{d}\tilde{\mathit{w}}=\Theta^3 \Xi_{ac}\;,\label{eq:conformal-rescaling-tildeXi}
\end{eqnarray}
\end{subequations}
where we have defined
\begin{eqnarray}
&&\mathfrak{d}_{abcd}\equiv \frac{1}{2} (d_{abcd} - i\ d^{*}{}_{abcd})\;,\\
&&\mathcal{D}_{abcd}\equiv \mathfrak{d}_{ab}{}^{pq}\mathfrak{d}_{pqcd} -  
\frac{\mathit{a}}{6}{\mathcal{G}}_{abcd} - \frac{\mathit{b}}{\mathit{a}}\mathfrak{d}_{abcd}\;,
\label{eq:define-mathcald}\\
&& \Xi_{ac}\equiv (\mathfrak{d}_{a}{}^{b}{}_{c}{}^{d}-\mathit{w}\mathcal{G}_{a}{}^{b}{}_{c}{}^{d})
\nabla_{b}\tilde{\mathit{w}} \nabla_{d}\tilde{\mathit{w}}.
\end{eqnarray}
\label{prop:rescaling-properties}
\end{proposition}
\proof This is a straightforward computation carried out by using the relations
(\ref{eq:unphysicaltophysical-downstairs}), (\ref{eq:unphysicaltophysical-upstairs}), 
(\ref{eq:etaunphystoetaphys}), (\ref{eq:weylunphystoweylphys}) and
(\ref{eq:rescaledweyl}).
\qed
\begin{remark}\em
We note that $\Xi_{ac}$ is a concomitant of both the unphysical metric $g_{ab}$ and the conformal factor $\Theta$.
It can be written in the following form
\begin{equation}
\Xi_{ac}=\Theta^6(\Xi^0)_{ac}+(\mathfrak{d}_{a}{}^{b}{}_{c}{}^{d}-\mathit{w}\mathcal{G}_{a}{}^{b}{}_{c}{}^{d})
(3\mathit{w}\Theta^5(\nabla_b\Theta\nabla_d\mathit{w}+\nabla_d\Theta\nabla_b\mathit{w})
+9\mathit{w}^2\Theta^4\nabla_b\Theta\nabla_d\Theta)\;,
\label{eq:xitoxi0}
\end{equation}
where
\begin{equation}
(\Xi^0)_{ac}\equiv(\mathfrak{d}_{a}{}^{b}{}_{c}{}^{d}-\mathit{w}\mathcal{G}_{a}{}^{b}{}_{c}{}^{d})
\nabla_{b}\mathit{w}\nabla_{d}\mathit{w}.
\end{equation}
The tensor $(\Xi^0)_{ac}$ is a concomitant of the unphysical metric $g_{ab}$ only.
\end{remark}

Next we introduce the rescaled Killing 1-form
\begin{equation}
\xi_a\equiv\Theta^2\tilde{\xi}_a.
\label{eq:rescaled-killing}
\end{equation}
By construction $\xi^a$ is a conformal Killing vector in the unphysical spacetime. Combining this with 
(\ref{eq:conformal-rescaling-tildeXi}),(\ref{eq:invariant-killing}) and (\ref{eq:conformal-rescaling-varphi})
we deduce
\begin{equation}
 \Xi_{ac}=\Theta^4\varphi\xi_a\xi_b.
\end{equation}

\begin{theorem}
At those points where $\Theta\neq 0$ 
the unphysical spacetime $(\mathcal{M},g_{ab})$ is conformal to a Petrov type D physical 
spacetime if and only if $\mathcal{D}_{abcd}=0$.
\label{theo:typeDunphysical}
\end{theorem}
\proof
This is a direct consequence of eq.(\ref{eq:conformal-rescaling-D}) and 
Theorem \ref{theo:type-D}.
\qed

\section{Construction of type D conformal initial data}
\label{sec:conformal-data-D}
We recall the standard construction of an initial data set for the physical vacuum Einstein equations  
$(\tilde{\mathcal{M}},\tilde{g}_{ab})$.
\begin{theorem} 
Let $(\tilde{\mathcal{S}},\tilde{h}_{ij})$ be a Riemannian manifold 
and define $\tilde{h}^{ij}$ as the inverse of 
$\tilde{h}_{ij}$
Suppose that there exists a symmetric tensor field $\tilde{K}_{ij}$ on $\tilde{\mathcal{S}}$ which satisfies
the conditions (vacuum constraints)
\begin{eqnarray}
&& \tilde{r}+ \tilde{K}^2-\tilde{h}^{ip}\tilde{h}^{jq}\tilde{K}_{pq}\tilde{K}_{ij}=2\lambda, \label{Hamiltonian}\\
&&\tilde{h}^{qp}\tilde{D}_p\tilde{K}_{iq}-\tilde{D}_i\tilde{K}=0, \label{Momentum}
\end{eqnarray}
where $\tilde{K}\equiv \tilde{h}^{ij}\tilde{K}_{ij}$ and $\tilde{D}_i$ is the covariant derivative compatible with $\tilde{h}_{ij}$. 
Provided that $\tilde{h}_{ij}$ and $\tilde{K}_{ij}$ are 
smooth, there exists an isometric embedding $\phi$ of
$\tilde{\mathcal{S}}$ into a globally hyperbolic, vacuum solution
$(\tilde{\mathcal{M}},\tilde{g}_{\mu\nu})$ of the Einstein field equations
\begin{equation}
\tilde{R}_{ab} = \lambda \tilde{g}_{ab}\;,\quad\lambda\in\mathbb{R}.
\end{equation}
The set
$(\tilde{\mathcal{S}},\tilde{h}_{ij}, \tilde{K}_{ij})$ is then called a vacuum initial data
set and the spacetime $(\tilde{\mathcal{M}},\tilde{g}_{\mu\nu})$ is the data
development. Furthermore the spacelike hypersurface
$\phi(\tilde{\mathcal{S}})$ is a Cauchy hypersurface in $(\tilde{\mathcal{M}},\tilde{g}_{\mu\nu})$.
\label{theorem:vacuum-data}
\end{theorem}

\subsection{Killing initial data equations and their conformal rescaling}
The following definition has been taken from \cite{CHRUSCIEL-BEIG-KID}.
\begin{definition}
Two tensor fields $\tilde{Y}$ and $\tilde{Y}_a$ fulfill the Killing Initial Data (KID)
conditions on $(\tilde{\mathcal{S}},\tilde{h}_{ab})$ if and only if:
\begin{subequations}
\begin{eqnarray}
&&
	-2\tilde{Y}\tilde{K}_{ab} +  
	\tilde{D}_{a}\tilde{Y}_{b} + 
	\tilde{D}_{b}\tilde{Y}_{a} = 0\;,\label{eq:kid-1}\\
&&
	\lambda\tilde{Y}\tilde{h}_{ab} - 
	(
		\tilde{h}^{cd} \tilde{K}_{ab} \tilde{K}_{cd} - 
		2\tilde{h}^{cd}\tilde{K}_{ac} \tilde{K}_{db} + 
		\tilde{r}_{ab}
	) \tilde{Y} + 
	\tilde{D}_{b}\tilde{D}_{a}\tilde{Y} -  
	\mathcal{L}_{\tilde{Y}} \tilde{K}_{ab} = 0.
\label{eq:kid-2}
\end{eqnarray}
\end{subequations}
\end{definition}
For us the relevance of the KID equations is that any pair of tensor fields 
$(\tilde{Y}, \tilde{Y}_a)$ solving them gives rise to a Killing vector $\tilde{\zeta}^a$ in the physical space-time 
$(\tilde{\mathcal M},\tilde{g}_{ab})$. 
The orthogonal splitting of $\tilde{\zeta}^a$ 
with respect to any $\tilde{\mathcal{S}}$-normal unit timelike vector field $\tilde{n}^a$ is 
\begin{equation}
\breve{\tilde{\zeta}}_a= \tilde{Y} \tilde{N}_a+ \tilde{Y}_a\;,\quad \tilde{Y}\equiv -\tilde{N}_a\tilde{\zeta}^a\;,\quad 
 \tilde{Y}_a \equiv\tilde{h}_{ab}\tilde{\zeta}^b\;,\quad \tilde{N}_a\equiv\tilde{g}_{ab}\tilde{n}^b\;,\quad
\breve{\tilde{\zeta}}_a\equiv\tilde{g}_{ab}\tilde{\zeta}^b.
\label{eq:killing-splitting}
\end{equation}
See \cite{CHRUSCIEL-BEIG-KID,COLL77,MONCRIEF-KID} for proofs of the above statements.
\begin{remark}\em
If the orthogonal splitting of the 1-form $\breve{\tilde{\zeta}}_a$ with respect to the physical metric is given
by
\begin{equation}
\breve{\tilde{\zeta}}_a= \tilde{Y} \tilde{N}_a+ \tilde{Y}_a\;,
\end{equation}
then using (\ref{eq:normal-unphys-to-normal-phys})-(\ref{eq:hKToTildehK}) we deduce that
its orthogonal splitting with respect to the unphysical metric can be written as
\begin{equation}
 \breve{\tilde{\zeta}}_a=\frac{Y}{\Theta^2}n_a+ \frac{Y_a}{\Theta^2}\;,
\label{eq:killing-physical-split}
\end{equation}
where
\begin{equation}
Y\equiv -n_a\tilde{\zeta}^a=\Theta \tilde{Y}\;,
\quad Y_a\equiv h_{ab}\tilde{\zeta}^b=\Theta^2 \tilde{Y}_a.
\label{eq:Killinglapseshift-phys-to-non-phys}
\end{equation}
When pull-backed to $\mathcal{S}$, these relations translate into the 
rescalings displayed at (\ref{eq:kid-rescaling}) below on those 
points of $\mathcal{S}$ where $\Theta$ is different from zero.
\label{rem:split-killing-phys}
\end{remark}

\begin{lemma}
Let $(Y,Y_a)$ be tensors on $\mathcal{S}$ and introduce the rescalings 
\begin{equation}
\tilde{Y} = \frac{Y}{\Omega}\;,\quad 
\tilde{Y}_{a} = \frac{Y_{a}}{\Omega^2}\;,
\label{eq:kid-rescaling}
\end{equation}
where $\Omega$ is a differentiable function different from zero on $\tilde{\mathcal{S}}\subset\mathcal{S}$. 
Then the tensors $Y$, $Y_a$ fulfill the following conditions on $\mathcal{S}$
\begin{subequations}
\begin{eqnarray}
&&
	\Omega (D_{a}Y_{b} + D_{b}Y_{a}) + 
	2 h_{ab} (\sigma Y -  \Omega Y^{c} D_{c}\Omega) = 
	2 \Omega K_{ab} Y\;,\label{eq:kid-rescaled-1}\\
&&
	2 \Omega^4 K_{a}{}^{c} K_{bc} Y-  
	r_{ab} \Omega^4 Y - 
	4 \sigma Y_{(b} D_ {a)}\Omega
	+ 2\big(2 \Omega Y^{c} D_ {(a}\Omega -  \Omega^2 D_{(a}Y^{c}\big)K_{b)c} 
	\nonumber \\
&& 
	+ 2\Omega \sigma D_{(a}Y_{b)} 
	+ \Omega K_{ab} 
	(
		- \Omega^3 Y K^{c}{}_{c} -  
		\Omega^2 \sigma Y + 
		Y^{c} D_ {c}\Omega
	)\nonumber \\
&& 
	- 2 \Omega^3 Y D_{b}D_ {a}\Omega+ 
	\Omega^4 D_{b}D_{a}Y - 
	\Omega^2 Y^{c} D_{c}K_{ab}\nonumber \\
&& 
	+ h_ {ab} \biggl(Y^{c} (\Omega D_ {c}\sigma -2 \sigma D_ {c}\Omega)+
	\Omega^3 (
		\sigma Y K^{c}{}_{c}- 
		(Y D_{c}D^{c}\Omega + D_{c}Y D^{c}\Omega)
	)\nonumber\\
&&
	+ \Omega^2 Y (\lambda-\sigma^2 + 
	3 D_ {c}\Omega D^{c}\Omega)\biggr)= 0\;,
\label{eq:kid-rescaled-2}
\end{eqnarray}
\end{subequations}
if and only if the rescaled tensors $\tilde{Y}$, $\tilde{Y}_{a}$ satisfy the KID conditions (\ref{eq:kid-1})-(\ref{eq:kid-2})
on $\tilde{\mathcal{S}}$ for $\tilde{h}_{ab}$, $\tilde{h}^{ab}$, $\tilde{K}_{ab}$ defined by (\ref{eq:TildehKTohK}).
\label{lemma:kid-rescaling}
\end{lemma}
\proof 
The relation between the Levi-Civita connections $\tilde{D}_a$ and $D_a$, compatible with the 
respective Riemannian metrics $\tilde{h}_{ab}$, $h_{ab}$, can be expressed in terms of the Christoffel 
tensor arising from the difference between the two connections:
\begin{equation}
\Gamma[D]^{e}{}_{ac}-\Gamma[\tilde{D}]^{e}{}_{ac} =\frac{1}{\Theta}(\delta_{c}{}^{e} D_{a}\Theta-h_{ca} h^{eb} D_{b}\Theta + \delta_{a}{}^{e} D_{c}\Theta).
\end{equation}
Using this tensor it is possible to express any covariant derivative with respect to $\tilde{D}_a$ in terms of $D_a$ and vice-versa.
Also the relations between the Ricci tensor of the connection $\tilde{D}_a$ and the Ricci tensor of 
the connection $D_a$ can be computed 
\begin{eqnarray}
&&\tilde{r}_{ac} = r_{ac} - \frac{2}{\Theta^2}h_{ac} D_{b}\Theta D^{b}\Theta 
 + \frac{1}{\Theta}(h_{ac} D_{b}D^{b}\Theta + D_{c}D_{a}\Theta)\;,\label{eq:r3-1}\\
&&r_{ac} = \tilde{r}_{ac} + \frac{2}{\Theta^2}\tilde{D}_{a}\Theta \tilde{D}_{c}\Theta 
-  \frac{1}{\Theta}(\tilde{D}_{c}\tilde{D}_{a}\Theta + \tilde{h}^{bd} \tilde{h}_{ac}\tilde{D}_{d}\tilde{D}_{b}\Theta).
\label{eq:r3-2}
\end{eqnarray}
Using the last of these relations together with (\ref{eq:hKToTildehK}) and (\ref{eq:kid-rescaling}) in 
(\ref{eq:kid-rescaled-1})-(\ref{eq:kid-rescaled-2}) leads us to 
(\ref{eq:kid-1})-(\ref{eq:kid-2}) after long algebra. Reciprocally, if we invert 
(\ref{eq:kid-rescaling}) to express $\tilde{Y}$, $\tilde{Y}_a$ in terms of 
$Y$, $Y_a$ and use (\ref{eq:TildehKTohK}) and (\ref{eq:r3-1}) in (\ref{eq:kid-1})-(\ref{eq:kid-2}) we 
get (\ref{eq:kid-rescaled-1})-(\ref{eq:kid-rescaled-2}).
\qed

\subsection{Vacuum conformal type D initial data}
The tensors $\mathfrak{d}_{abcd}$ and $\mathcal{D}_{abcd}$ are complex self-dual {\em Weyl candidates}. This means that they have the same 
algebraic properties as the Weyl tensor and 
this makes it possible to obtain their orthogonal splitting from a general formula involving the electric part of 
the corresponding Weyl candidate (see e.g. \cite{AGPTYPEDDATA})
\begin{eqnarray}
&&\mathfrak{d}_{abcd} = \mathcal{E}_{bd} (h_{ac} + n_{a} n_{c}) - \mathcal{E}_{ad}(h_{bc} + n_{b} n_{c}) 
+\mathcal{E}_{ac} (h_{bd} + n_{b} n_{d}) - \mathcal{E}_{bc} (h_{ad} + n_{a} n_{d})-\nonumber\\
&&\eta_{cdep} \mathcal{E}^{*}{}_{b}{}^{p} n_{a} n^{e} 
+ \eta_{cdep}\mathcal{E}^{*}{}_{a}{}^{p} n_{b} n^{e} -  
\eta_{abep}\mathcal{E}^{*}{}_{d}{}^{p} n_{c} n^{e} + 
\eta_{abep}\mathcal{E}^{*}{}_{c}{}^{p} n_{d} n^{e}\;,\label{eq:mathfrakd-split}\\
&& \mathcal{D}_{abcd} = \mathfrak{a}_{bd} (h_{ac} + n_{a} n_{c}) - \mathfrak{a}_{ad}(h_{bc} + n_{b} n_{c}) 
+\mathfrak{a}_{ac} (h_{bd} + n_{b} n_{d}) - \mathfrak{a}_{bc} (h_{ad} + n_{a} n_{d})-\nonumber\\
&&\eta_{cdep} \mathfrak{a}^{*}{}_{b}{}^{p} n_{a} n^{e} 
+ \eta_{cdep}\mathfrak{a}^{*}{}_{a}{}^{p} n_{b} n^{e} -  
\eta_{abep}\mathfrak{a}^{*}{}_{d}{}^{p} n_{c} n^{e} + 
\eta_{abep}\mathfrak{a}^{*}{}_{c}{}^{p} n_{d} n^{e}.
\label{eq:mathcal-split}
\end{eqnarray}
Where we have defined
\begin{eqnarray}
&&\mathcal{E}_{ab}\equiv\mathfrak{d}_{apbq}n^pn^q=\frac{1}{2}(d_{ab}-i\;d^*_{ab})\;,\label{eq:define-mcale}\\
&&\mathfrak{a}_{ab}\equiv\mathcal{D}_{apbq}n^pn^q.\label{eq:define-mathfraka}
\end{eqnarray}
Since the scalars $a$, $b$ and $w$ are defined from $\mathfrak{d}_{apbq}$ they can be rendered 
in terms of $\mathcal{E}_{ab}$
\begin{eqnarray}
&&\mathit{a} = 16 \mathcal{E}_{ab} \mathcal{E}^{ab}\;,\label{eq:split-a}\\
&&\mathit{b} = -64\mathcal{E}_{a}{}^{c} \mathcal{E}^{ab} \mathcal{E}_{bc}\;,\label{eq:split-b}\\
&&\mathit{w}\equiv-\frac{\mathit{b}}{2\mathit{a}}.\label{eq:split-w}
\end{eqnarray}
Using eqs. (\ref{eq:fund-quantity-4}) and (\ref{eq:fund-quantity-6}) in (\ref{eq:define-mcale})
we can find the expression of all these scalars in terms of quantities intrinsic to the 
initial data hypersurface.

\begin{proposition}
\begin{equation}
\mathcal{D}_{abcd}|_{\mathcal{S}}=0\Longleftrightarrow \mathfrak{a}_{ah}=0\;, 
\label{eq:dsigma0}
\end{equation}
where
\begin{equation}
\mathfrak{a}_{ah}=\frac{\mathit{a}}{12} h_{ah} - \frac{\mathit{b}}{\mathit{a}}\mathcal{E}_{ah} - 4  \mathcal{E}_{a}{}^{b} \mathcal{E}_{hb}.
\label{eq:define-a-mathfrak}
\end{equation}
\label{prop:D-initial}
\end{proposition}
\proof
Equation (\ref{eq:dsigma0}) is a direct consequence of
equation (\ref{eq:mathcal-split}) whereas (\ref{eq:define-a-mathfrak}) results from 
inserting the splitting of $\mathfrak{d}_{abcd}$ given by (\ref{eq:mathfrakd-split}) into
the definition of $\mathcal{D}_{abcd}$ stated by (\ref{eq:define-mathcald}) and then combining the result with
(\ref{eq:mathcal-split}).
\qed

We define now the following quantities, which can be regarded as, respectively, a scalar and a tensor defined on $\mathcal{S}$
\begin{eqnarray}
&&\mathit{w}^{\bot}{}_{a} \equiv \frac{\mathit{b} D_{a}\mathit{a} - \mathit{a} D_{a}\mathit{b}}{2 \mathit{a}^2}\;,\\
&&\mathit{w}^{\parallel}{} \equiv - \frac{6 K^{bc}}{\mathit{a}^3}\bigl(\mathit{b}^2 \
\mathcal{E}_{bc} + \mathit{a} (\mathit{b} \mathfrak{a}_{bc} - 12 \
\mathit{a} \mathcal{E}_{b}{}^{d} \mathfrak{a}_{cd})\bigr) -  \frac{\mathit{a} \mathit{b} K^{b}{}_{b} \
 - 16i \varepsilon_{cde} (\mathit{b} \
\mathcal{E}^{bc} + 3 \mathit{a} \mathfrak{a}^{bc}) \
D^{e}\mathcal{E}_{b}{}^{d}}{2 \mathit{a}^2}\;,\nonumber\\
&&
\end{eqnarray}
where
\begin{equation}
\varepsilon_{abc}\equiv\eta_{dabc}n^d.\\
\end{equation}
\begin{lemma}
\begin{equation}
\nabla_aw=n_a \mathit{w}^{\parallel}{}+\mathit{w}^{\bot}{}_a.
\label{eq:decompose-nabla-w}
\end{equation}
\label{lem:decompose-nabla-w}
\end{lemma}
\proof
From (\ref{eq:split-a})-(\ref{eq:split-w}) we deduce that $w$ can be rendered excusively in terms of 
scalars formed with $\mathcal{E}_{ab}$. Therefore 
to compute the orthogonal splitting of $\nabla_aw$ we need to compute first the orthogonal splitting of 
$\nabla_{a}\mathcal{E}_{bc}$. The latter turns out to be
\begin{equation}
\nabla_{a}\mathcal{E}_{bc} = (2K_{(c}{}^{d} n_{a)} + A^{d} n_{a} n_{c})
\mathcal{E}_{bd} + (2 K_{(b}{}^{d} n_{a)} + A^{d} n_{a} n_{b})\mathcal{E}_{cd} +
D_{a}\mathcal{E}_{bc} -  n_{a} \mathcal{L}_n \mathcal{E}_{bc}\;,
\label{eq:nabla-e}
\end{equation}
where $A^a\equiv n^a\nabla_a n^b$. The last term of (\ref{eq:nabla-e}) can be further worked out using  
the orthogonal splitting of (\ref{eq:cfe4}) which decomposes into the standard evolution and constraint equations
\begin{eqnarray}
&&\mathcal{L}_{\vec{\boldsymbol n}} \mathcal{E}_{cp} = -2 K^{b}{}_{b} \mathcal{E}_{cp} + \
2i a^{b}\mathcal{E}_{(c}{}^{d}\varepsilon_{p)bd} 
- i \epsilon_{(c}{}^{bd}D_{|b|}\mathcal{E}_{p)d} -  h_{cp}K^{bd} \mathcal{E}_{bd} + 5 \
K_{(c}{}^{b}\mathcal{E}_{p)b}\;,\\
&&D_{b}\mathcal{E}_{a}{}^{b} = -i \varepsilon_{acd} K^{bc} \mathcal{E}_{b}{}^{d}.
\end{eqnarray}
Using these results in $\nabla_a w$, eq. (\ref{eq:decompose-nabla-w}) follows after some manipulations.\qed

\begin{proposition}
If $\tilde{\eta}_a$ is a covector in $\tilde{\mathcal{M}}$ defined on an open set containing 
$\tilde{\mathcal{S}}$ such that its orthogonal splitting with respect to $n_a$ and $h_{ab}$ is given by
\begin{equation}
\tilde{\eta}_a=\frac{Y}{\Theta^2} n_a+\frac{Y_a}{\Theta^2}.
\label{eq:ot-eta}
\end{equation}
then
\begin{eqnarray}
&&(\tilde{\Xi}_{ac}-\tilde{\varphi}\tilde{\eta}_{a} \tilde{\eta}_{c})|_{\tilde{S}}=0
\Longleftrightarrow\label{eq:killing-candidate-computation}\\
&&Y_a Y_b=\frac{Q_{ab}}{\Omega^4}\;,
\label{eq:restriction-3}
\end{eqnarray} 
where in the previous equation, $Y$, $Y_a$ and $Q_{ab}$ are understood 
as quantities defined on $\mathcal{S}$ through the relations
\begin{eqnarray}
&& Y^2 = \frac{1}{2\varphi}\bigg(\mathit{w} \bigl(3 \sigma^{b} \mathit{w} (6 \sigma^{d} \
\mathcal{E}_{bd} + 3 \sigma_{b} \mathit{w} + 2 \Omega \
\mathit{w}^{\bot}{}_{b}) + \Omega^2 \mathit{w}^{\bot}{}_{b} \
\mathit{w}^{\bot}{}^{b}\bigr) + \nonumber\\
&& 2 \Omega \mathcal{E}_{bd} (6 \
\sigma^{b} \mathit{w} + \Omega \mathit{w}^{\bot}{}^{b}) \
\mathit{w}^{\bot}{}^{d}\bigg)\;,\label{eq:Killing-lapse}\\
&& Y_{c} \equiv - \frac{\Omega}{2Y\varphi}\bigg((3 \sigma \mathit{w} -  \Omega \
\mathit{w}^{\parallel}{}) (6 \sigma^{b} \mathcal{E}_{cb} \mathit{w} + \
3 \sigma_{c} \mathit{w}^2 + 2 \Omega \mathcal{E}_{cb} \
\mathit{w}^{\bot}{}^{b} + \Omega \mathit{w} \mathit{w}^{\bot}{}_{c}) + \nonumber\\
&& 6i \Omega \varepsilon_{cba} \sigma^{b} \mathcal{E}_{d}{}^{a} \
\mathit{w} \mathit{w}^{\bot}{}^{d} + 2i \varepsilon_{cda} \
\mathcal{E}_{b}{}^{a} \bigl(\Omega^2 \mathit{w}^{\bot}{}^{b} \
\mathit{w}^{\bot}{}^{d} + 3 \sigma^{b} \mathit{w} (3 \sigma^{d} \
\mathit{w} + \Omega \mathit{w}^{\bot}{}^{d})\bigr)\bigg)\;,\label{eq:Killing-shift}\\
&&\varphi Q_{ac} = \frac{1}{2} \Omega^4 (2 \mathcal{E}_{ac} + h_{ac} \mathit{w}) (-3 \
\sigma \mathit{w} + \Omega \mathit{w}^{\parallel}{})^2 -\nonumber \\
&& i \Omega^4 (\varepsilon_{cbd} \mathcal{E}_{a}{}^{d} + \varepsilon_{abd} \
\mathcal{E}_{c}{}^{d}) (-3 \sigma \mathit{w} + \Omega \
\mathit{w}^{\parallel}{}) (3 \sigma^{b} \mathit{w} + \Omega \
\mathit{w}^{\bot}{}^{b}) + \nonumber\\
&&\bigl(h^{bd} \mathcal{E}_{ac} -  \
h^{b}{}_{c} \mathcal{E}_{a}{}^{d} -  h_{a}{}^{d} \
\mathcal{E}^{b}{}_{c} + \frac{1}{2} h_{a}{}^{d} h^{b}{}_{c} \mathit{w} + h_{ac} (\mathcal{E}^{bd} - 
\frac{1}{2} h^{bd}\mathit{w})\bigr)\times\nonumber\\ 
&&(3 \Omega^2 \sigma_{b} \mathit{w} + \Omega^3 \mathit{w}^{\bot}{}_{b}) 
(3 \Omega^2 \sigma_{d} \mathit{w} + \Omega^3 \mathit{w}^{\bot}{}_{d})\label{eq:Killinglapseshift}.
\end{eqnarray}
\label{prop:q-splitting}
\end{proposition}
\proof 
To find out the conditions arising from (\ref{eq:killing-candidate-computation}) 
we need to find the orthogonal splitting of the tensor
\begin{equation}
\tilde{T}_{ab}\equiv\tilde{\Xi}_{ac}-\tilde{\varphi}\tilde{\eta}_{a} \tilde{\eta}_{c}\;,
\label{eq:def-tilde-T}
\end{equation}
and set each of its spatial parts to zero.
This is a straightforward albeit tedious computation that requires the following steps:
\begin{itemize}
 \item computation of the orthogonal splitting of $\tilde{\eta}_a$
\begin{equation}
\tilde{\eta}_a=\tilde{U}n_a+\tilde{U}_a.
\end{equation}
\item Computation of the orthogonal splitting of $\nabla_aw$ (see lemma \ref{lem:decompose-nabla-w})
\begin{equation}
\nabla_aw=n_a \mathit{w}^{\parallel}{}+\mathit{w}^{\bot}{}_a.
\end{equation}
\item The restriction of eq. (\ref{eq:conformal-rescaling-varphi}) to $\mathcal{S}$
\begin{equation}
\tilde{\varphi}|_{\mathcal{S}} = \Omega^{11} \varphi|_{\mathcal{S}}.
\end{equation}
\item Eq. (\ref{eq:cf-data5}) 
$$
\nabla_{a}\Theta = - n_{a} \sigma + \sigma_{a}
$$
\item Eq. (\ref{eq:mathfrakd-split})

\end{itemize}
Next one uses eq. (\ref{eq:xitoxi0}) on eq. (\ref{eq:conformal-rescaling-tildeXi}) and 
carries out in the resulting expression the steps described above. This yields the orthogonal splitting of 
$\tilde{\Xi}_{ab}$ which is then used in (\ref{eq:def-tilde-T}) to find the orthogonal splitting of
$\tilde{T}_{ab}$. We write such orthogonal splitting in the form
\begin{equation}
 \tilde{T}_{ab}=A n_a n_b+B_{(a}n_{b)}+C_{ab}\;,
\end{equation}
where $A$, $B_a$, $C_{ab}$ are spatial and known. Thus $\tilde{T}_{ab}|_{\tilde{\mathcal{S}}}=0$ 
if and only if $A=0$, $B_a=0$ and $C_{ab}=0$.

\begin{itemize}
 \item The condition $A=0$ corresponds to (\ref{eq:Killing-lapse}) if we set
\begin{equation}
 \tilde{U}=\frac{Y}{\Omega^2}.\label{eq:lapse-tildelapse}
\end{equation}
\item The condition $B_a=0$ corresponds to (\ref{eq:Killing-shift}) if we set
\begin{equation}
\tilde{U}_a=\frac{Y_a}{\Omega^2}.\label{eq:shift-tildeshift}
\end{equation}
\item The condition $C_{ab}=0$ corresponds to (\ref{eq:Killinglapseshift}).
\end{itemize}
\qed

\subsection{The main results}
\begin{figure}[t]
\centering
\includegraphics[width=\textwidth,keepaspectratio=true]{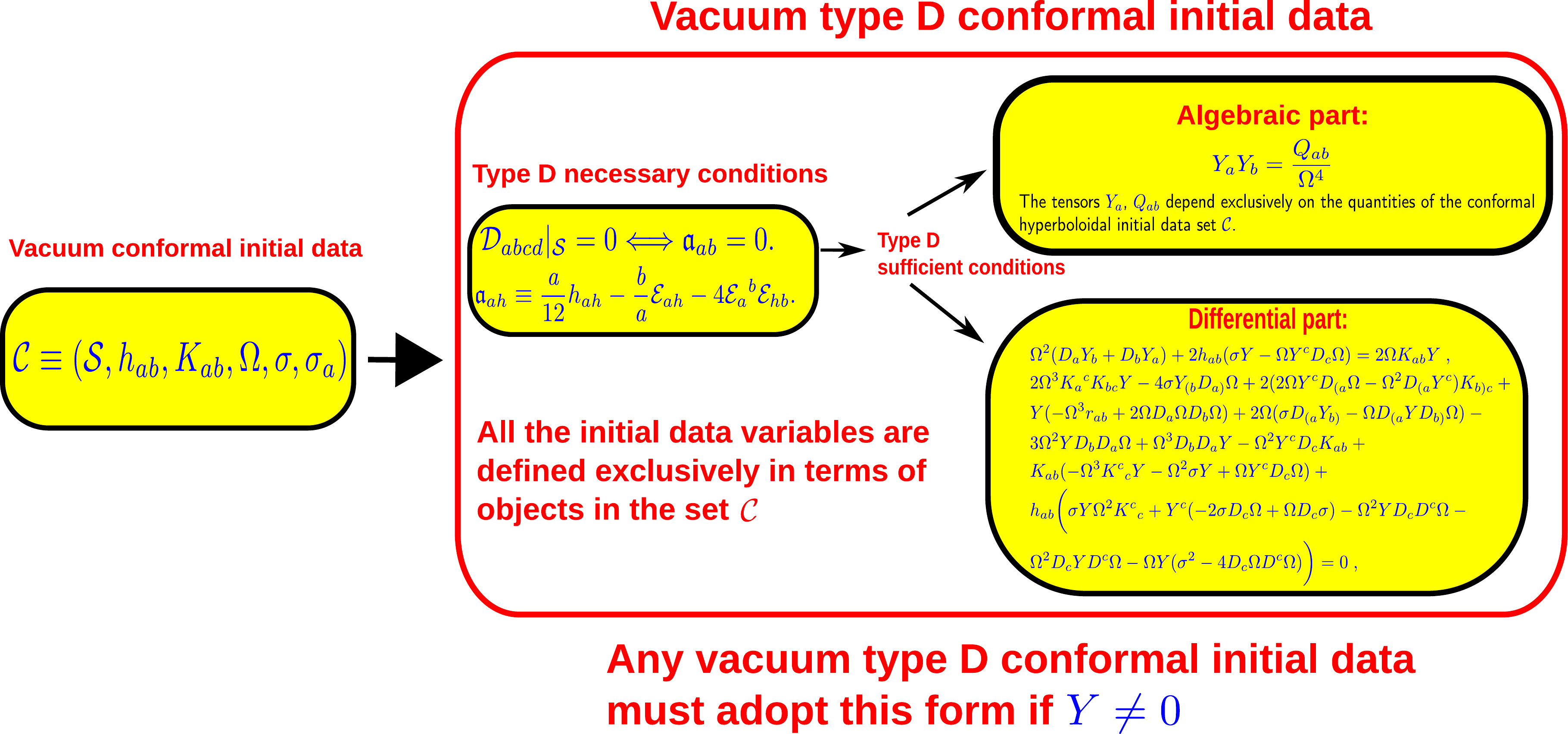}
% MainResult.pdf: 0x0 px, 300dpi, 0.00x0.00 cm, bb=
\caption{Summary of the vacuum type 
D conformal initial data characterization 
found in theorems \ref{theo:conformal-petrov-d} and \ref{theo:conformal-petrov-d-necessary}. 
\label{fig:main-result}}
\end{figure}

We present next theorems \ref{theo:conformal-petrov-d} and 
\ref{theo:conformal-petrov-d-necessary} which are
the main results of the paper (see \ref{fig:main-result} for
a graphical depiction of these results).

\begin{theorem}
Let $(\tilde{\mathcal{S}}, \tilde{h}_{ab},\tilde{K}_{ab})$ be a smooth $\lambda$-vacuum initial data set and consider 
a smooth conformal initial data set constructed from it (see Definition \ref{def:conformal-initial-data}) 
\begin{equation}
{\mathcal C}\equiv(\mathcal{S},h_{ab},K_{ab},\Omega,\sigma,\sigma_a)\;,
\label{eq:conf-initial-data}
\end{equation}
fulfilling the conformal Hamiltonian and momentum constraints 
(\ref{eq:conf-hamiltonian-constraint})-(\ref{eq:conf-momentum-constrant}), where $\tilde{\mathcal{S}}\subset\mathcal{S}$. 
Use the data of $\mathcal{C}$ to define on $\mathcal{S}$ the quantities $s$, 
$\theta_a$, $\theta_{ab}$, $d_{ab}$, $d^*_{ab}$, $d_{abc}$
according to the prescriptions laid by
(\ref{eq:fund-quantity-1})-(\ref{eq:fund-quantity-6}).
From these quantities, we define on $\mathcal{S}$ the tensors
$\mathcal{E}_{ab}$, 
$Y$, $Y_a$ using resp. \eqref{eq:define-mcale}, 
(\ref{eq:Killing-lapse})-(\ref{eq:Killing-shift}). 
Assume further that on $\mathcal{S}$
\begin{enumerate}
\item $\mathcal{E}_{ab}$ , $h_{ab}$, are subject to the algebraic
condition (see (\ref{eq:define-a-mathfrak}))
\begin{equation}
\mathfrak{a}_{ab}=0 \;, 
\label{eq:type-D-initial}
\end{equation}

\item $Y$, $Y_a$ are subject to the algebraic condition
\eqref{eq:restriction-3}
\begin{equation}
 Y_a Y_b=\frac{Q_{ab}}{\Omega^4}\;,
\label{eq:restriction-3-bis}
\end{equation}
and to the differential conditions 
(\ref{eq:kid-rescaled-1})-(\ref{eq:kid-rescaled-2}) on $\mathcal{S}$.
Moreover, $Y\neq 0$ on $\tilde{\mathcal{S}}$.
\end{enumerate}
Then, there is an open subset contained in the data development of $\mathcal{C}$ where $(\mathcal{M},g_{ab})$ 
is conformally related to 
a $\lambda$-vacuum Petrov type D solution of the Einstein field equations corresponding 
to the data development of $(\tilde{\mathcal{S}}, \tilde{h}_{ab},\tilde{K}_{ab})$.
\label{theo:conformal-petrov-d}
\end{theorem}
\proof 
The fact that the data $\mathcal{C}$ is a solution of the conformal Hamiltonian and momentum constraints implies according to 
Theorems \ref{theo:conf-constraints} and \ref{theo:conf-ham-mom} that a solution of the conformal equations 
(\ref{eq:cfe0})-(\ref{eq:cfe4}) exists such that $(\mathcal{M},g_{ab})$ is conformal to a vacuum solution 
$(\tilde{\mathcal{M}},\tilde{g}_{ab})$ of the Einstein's equations (see Proposition \ref{prop:conformal-solution}). 
Furthermore Theorem
\ref{theo:conf-constraints} tells us that $(\tilde{\mathcal{M}},\tilde{g}_{ab})$ arises from 
the vacuum initial data $(\tilde{\mathcal{S}}, \tilde{h}_{ab},\tilde{K}_{ab})$.

Now, the condition $\mathfrak{a}_{ab}=0$ on $\mathcal{S}$ entails, via Proposition \ref{prop:D-initial} 
that $\mathcal{D}_{abcd}|_{\mathcal{S}}=0$ which, by (\ref{eq:conformal-rescaling-D}), leads to $\tilde{\mathcal{D}}_{abcd}|_{\tilde{\mathcal{S}}}=0$.
Thus, it remains to show that $\tilde{\mathcal{D}}_{abcd}=0$ on an open subset of $(\tilde{\mathcal{M}},\tilde{g}_{ab})$ 
contained in the data development of $\mathcal{C}$.
To that end we use (\ref{eq:kid-rescaling}) to introduce the quantities 
$\tilde{Y}$, $\tilde{Y}_a$ on $\tilde{\mathcal{S}}$. 
The differential conditions (\ref{eq:kid-rescaled-1})-(\ref{eq:kid-rescaled-2}) imply, according to 
Lemma \ref{lemma:kid-rescaling} that $\tilde{Y}$, $\tilde{Y}_a$ fulfill the KID conditions 
(\ref{eq:kid-1})-(\ref{eq:kid-2}) and thus 
there exists a Killing 1-form ${\tilde\xi}_a$ in $(\tilde{\mathcal{M}},\tilde{g}_{ab})$
with the properties displayed by 
(\ref{eq:killing-splitting}). Now, Remark \ref{rem:split-killing-phys} tells us that 
this Killing 1-form fulfills
condition (\ref{eq:ot-eta}) in some neighbourhood of $\mathcal{S}$
(therefore a neighbourhood of $\tilde{\mathcal S}$ as $\tilde{\mathcal S}\subset{\mathcal S}$)
and since, by assumption, (\ref{eq:Killing-lapse})-(\ref{eq:Killing-shift})
and (\ref{eq:restriction-3}) are fulfilled on $\mathcal{S}$, we can apply Proposition \ref{prop:q-splitting} taking 
as the covector $\tilde{\eta}_a$ the Killing 1- form ${\tilde{\xi}}_a$
and conclude that
\begin{equation}
(\tilde{\Xi}_{ac}-\tilde{\varphi}\tilde{\xi}_{a} \tilde{\xi}_{c})|_{\tilde{\mathcal{S}}}=0.
\label{eq:initial-data1}
\end{equation}
The proof is now similar to that of Theorem 6 in \cite{AGPTYPEDDATA} but we provide here the details 
for the sake of completeness: the Killing property
of $\tilde{\xi}_a$ automatically yields
\begin{equation}
\pounds_{\tilde{\boldsymbol\xi}}\tilde{D}_{abcd}=0\;,\quad
\pounds_{\tilde{\boldsymbol\xi}}(\tilde{\Xi}_{ac}-\tilde{\varphi}\tilde{\xi}_{a} \tilde{\xi}_{c})=0.
\label{eq:propagation-system}
\end{equation}
These equations can be regarded as a linear system for the variables $\tilde{D}_{abcd}$ and $\tilde{\Xi}_{ac}-\tilde{\varphi}\tilde{\xi}_{a} \tilde{\xi}_{c}$
with initial data given by (\ref{eq:initial-data1}) and $\tilde{\mathcal{D}}_{abcd}|_{\tilde{\mathcal{S}}}=0$.
The data of the system are trivial and non-characteristic given that $\tilde{Y}\neq 0$
(the characteristic points of the system are those in which $\tilde{\xi}_a$ is tangent to $\tilde{\mathcal{S}}$).
Hence we conclude that there is an open subset $\mathcal{U}\subset \tilde{\mathcal{M}}$ containing $\tilde{\mathcal{S}}$
where one has 
\begin{equation}
 \mathcal{\tilde{D}}_{abcd}=0\;,\quad 
 \tilde{\Xi}_{ac}-\tilde{\varphi}\tilde{\xi}_{a} \tilde{\xi}_{c}=0\;,\quad
 \tilde{\nabla}_{a}\tilde{\xi}_b + \tilde{\nabla}_{b}\tilde{\xi}_a = 0.
\label{eq:type-D-development}
\end{equation}
The second and third equations are actually redundant and can be dropped 
(see \cite{AGPTYPEDDATA} for more details\footnote{The reasoning of \cite{AGPTYPEDDATA} was formulated 
for the case with $\lambda=0$ but it still holds 
when $\lambda\neq 0$}.). 
Thus $(\tilde{\mathcal{M}},\tilde{g}_{ab})$ is of Petrov
type D in the open set $\mathcal{U}$ that is contained in the data development of 
$\mathcal{C}$ as $\tilde{\mathcal{M}}\subset\mathcal{M}$.
\qed

Theorem \ref{theo:conformal-petrov-d-necessary} admits a converse that is formulated next
\begin{theorem}
The initial data of any solution of the conformal equations (\ref{eq:cfe0})-(\ref{eq:cfe4})
conformal to a vacuum type D solution of the Einstein equations must comply with 
points 1 and 2 of Theorem \ref{theo:conformal-petrov-d}.
\label{theo:conformal-petrov-d-necessary}
\end{theorem}
\proof To prove this theorem, let us suppose that we have a solution $(\mathcal{M}, g_{ab})$ of the conformal 
equations (\ref{eq:cfe0})-(\ref{eq:cfe4}) arising from an initial data set as described by (\ref{eq:conf-initial-data}).
By assumption this solution is conformal to a vacuum type D solution of the Einstein field equations $(\tilde{\mathcal M},\tilde{g}_{ab})$
according to the relation (\ref{eq:unphysicaltophysical-downstairs}).
This implies, according to Theorem \ref{theo:type-D} that $\tilde{\mathcal{D}}_{abcd}$ vanishes on the physical 
space-time $(\tilde{\mathcal{M}},\tilde{g}_{ab})$ and hence from (\ref{eq:conformal-rescaling-D}) we have that 
$\mathcal{D}_{abcd}=0$ in the un-physical space-time $(\mathcal{M},g_{ab})$ whenever $\Theta\neq 0$. 
Combining this with Proposition \ref{prop:D-initial}
leads to point 1 of Theorem  \ref{theo:conformal-petrov-d}. To show that
point 2 holds too, we appeal to Proposition \ref{prop:killing-candidate} to deduce the existence of a Killing 1-form 
$\tilde{\xi}_a$ which according to Remark \ref{rem:split-killing-phys} has the following orthogonal splittings
in the physical and the unphysical space-times
\begin{equation}
\tilde{\xi}_a= \tilde{Y} \tilde{N}_a+ \tilde{Y}_a\;,\quad
\tilde{\xi}_a=\frac{Y}{\Theta^2}n_a+ \frac{Y_a}{\Theta^2}.
\end{equation}
Since on $\mathcal{S}$, the variables $\tilde{Y}$, $\tilde{Y}_a$, $Y$, $Y_a$ are related in the way shown by 
(\ref{eq:Killinglapseshift-phys-to-non-phys}) and $\tilde{Y}$, $\tilde{Y}_a$ fulfill (\ref{eq:kid-1})-(\ref{eq:kid-2})
due to the fact that $\tilde{\xi}_a$ is a Killing 1-form, 
then, by Lemma \ref{lemma:kid-rescaling},  $Y$, $Y_a$ fulfill (\ref{eq:kid-rescaled-1})-(\ref{eq:kid-rescaled-2}) on $\mathcal{S}$.
Moreover, from Proposition \ref{prop:q-splitting} we find that $Y$, $Y_a$ have on $\mathcal{S}$ 
the values given by (\ref{eq:Killing-lapse})-(\ref{eq:Killinglapseshift}). Therefore, combining these results, 
we finally conclude that point 2 also holds.
\qed

Theorems \ref{theo:conformal-petrov-d} and  
\ref{theo:conformal-petrov-d-necessary}
provide necessary and sufficient conditions 
which must be satisfied by a conformal initial data set $\mathcal{C}$ 
of the conformal equations in order that the unphysical space-time $(\mathcal{M}, g_{ab})$ be conformal to a vacuum type D solution 
of the Einstein's field equations. This result does not state anything about the existence of actual data fulfilling the 
given conditions and this is in fact an independent open problem (see section \ref{sec:conformal-limit} for more details). 
Compare this with the similar problem of the 
generic existence of hyperboloidal data for the conformal equations \cite{ANDERSSON1994,ANDERSSON1993,ANDERSSON1992,JANOS96}.

\section{Conformal initial data for the Kerr solution}
\label{sec:kerr-conformal-data}
The following result was proven in \cite{FERSAEZKERR} but we adopt the formulation presented in \cite{AGPTYPEDDATA,AGPTYPEDDATA-ERRATUM}.
\begin{theorem}
Under the conditions of Theorem \ref{theo:type-D}
a vacuum ($\lambda$=0) space-time $(\tilde{\mathcal{M}},\tilde{g}_{ab})$ is locally isometric to the Kerr 
solution with non-vanishing mass (non-trivial Kerr solution) if and only if the following additional conditions hold
\begin{eqnarray}
&&\tilde{\xi}_{[a}\tilde{\xi}^*_{b]}=0\Longleftrightarrow \tilde{\Xi}_{a[b}\tilde{\Xi}^*_{c]d}=0\;,
\label{eq:killing-property-kerr}\\
&&\mbox{\em Im}(\tilde{Z}^{3}(\tilde{w}^*)^8)=0\;,\quad \tilde{Z}\equiv 
\tilde{g}^{ab}\tilde{\nabla}_a \tilde{w}\tilde{\nabla}_b\tilde{w}\;,
\label{eq:nutzero}\\
&&
\frac{\mbox{\em Re}(\tilde{Z}^{3}(\tilde{w}^*)^8)}{\big(18 \mbox{\em Re}\big(\tilde{w}^3\tilde{Z}^*\big)
-|\tilde{Z}|^2\big)^3}<0,\;
(\mbox{if}\ 18 \mbox{\em Re}\big(\tilde{w}^3\tilde{Z}^*\big)-|\tilde{Z}|^2\neq 0),\;
\label{eq:epsilongeq0}\\
&&
\mbox{\em Re}(\tilde{Z}^{3}(\tilde{w}^*)^8)=0\Longleftrightarrow \tilde{g}^{ab}\tilde{\xi}_a\tilde{\xi}^*_b=0,\;
(\mbox{if}\
18 \mbox{\em Re}\big(\tilde{w}^3\tilde{Z}^*\big)-|\tilde{Z}|^2= 0)\;,
\label{eq:epsilongeq00}
\end{eqnarray}
where $\tilde{\xi}_{a}$ is defined by (\ref{eq:invariant-killing}).
\label{theo:kerr-local}
\end{theorem}

\begin{theorem} There exists an open subset of the unphysical spacetime $(\mathcal{M},g_{ab})$ that is locally conformal 
to the non-trivial Kerr solution if and only if
\begin{subequations}
\begin{eqnarray}
&&\mathcal{D}_{abcd}=0\;,\\
&&{\xi}_{[a}{\xi}^*_{b]}=0\Longleftrightarrow \Xi_{a[b}\Xi^*_{c]d}=0\;,
\label{eq:killing-property-cfkerr}\\
&&\mbox{\em Im}\big(Z^3(w^*)^8\big)=0\;,\quad Z\equiv g^{ab}\lambda_a\lambda_b\;,\quad
\lambda_a\equiv 3w\nabla_a\Theta+\Theta\nabla_a w\;, 
\label{eq:nut-unphysical-zero}\\
&&\frac{\mbox{\em Re}\big(Z^3(w^*)^8\big)}{\bigg(18\Theta^{3}\mbox{\em Re}(w^3 Z^*)-|Z|^2\bigg)^3}<0\;,\quad
\mbox{if}\;\;\;\; 18\Theta^{3}\mbox{\em Re}(w^3 Z^*)-|Z|^2\neq 0\;,\label{eq:positive-mass-unphysical}\\
&& \mbox{\em Re}(w^3 Z^*)=0\Longleftrightarrow g^{ab}\xi_a\xi^*_b=0\;,\quad \mbox{if}\;\;\; 18\Theta^{3}\mbox{\em Re}(w^3 Z^*)-|Z|^2=0\;,
\label{eq:positive-mass-unphysical-2}
\end{eqnarray}
\end{subequations}
where $\xi^a$ is defined by (\ref{eq:rescaled-killing}).
\label{theo:kerr-conformal}
\end{theorem}

\proof This is a straightforward computation involving the relations found in Proposition \ref{prop:rescaling-properties} and 
their replacement in (\ref{eq:killing-property-kerr})-(\ref{eq:epsilongeq00}).
\qed

\begin{theorem}
Under the hypotheses of Theorem \ref{theo:conformal-petrov-d} and whenever the conformal factor 
$\Omega$ does not vanish, the initial data set $\mathcal{C}$ are data for a spacetime conformal to the 
non-trivial Kerr solution (unphysical Kerr spacetime) if and only if 
\begin{equation}
\mbox{\em Im}(YY^*_b)=0\;,\quad Y_{[a}Y^*_{b]}=0\;,
\label{eq:kerr-conformal-initial-data}
\end{equation}
and conditions (\ref{eq:nut-unphysical-zero})-(\ref{eq:positive-mass-unphysical-2}) of Theorem \ref{theo:kerr-conformal} 
hold replacing $\Theta$ by $\Omega$ and with the following definition for $Z$:
\begin{equation}
Z\equiv-(\Omega w^{\parallel}-3\sigma w)^2+
(3w\sigma_a+\Omega\mathit{w}^{\bot}_a)(3w\sigma^a+\Omega(\mathit{w}^{\bot}){}^a)\;,\label{eq:Z-split}\\
\end{equation}
where the quantities
\begin{eqnarray}
&&\mathit{a} = 16 \mathcal{E}_{ab} \mathcal{E}^{ab}\;,\\
&&\mathit{b} = -64\mathcal{E}_{a}{}^{c} \mathcal{E}^{ab} \mathcal{E}_{bc}\;,\\
&&\mathit{w}=-\frac{\mathit{b}}{2\mathit{a}}\;,
\end{eqnarray}
are now understood as defined from the conformal initial data set $\mathcal{C}$ using (\ref{eq:define-mcale}), 
(\ref{eq:fund-quantity-4}) and (\ref{eq:fund-quantity-6}). 
\label{theo:kerr-conformal-data}
\end{theorem}
\proof
Use (\ref{eq:killing-physical-split}) and (\ref{eq:rescaled-killing}) to obtain
\begin{equation}
\xi_a=Y n_a+Y_a.
\end{equation}
The combination of this with (\ref{eq:killing-property-cfkerr}) leads immediately to the conditions
\begin{equation}
\mbox{Im}(Y Y^*_b)=0\;,\quad Y_{[a}Y^*_{b]}=0.
\end{equation}
Moreover, the variables $Y$, $Y_a$ have the values given by (\ref{eq:Killing-lapse})-(\ref{eq:Killing-shift}) 
on the initial data hypersurface $\mathcal{S}$ as defined in Theorem \ref{theo:conformal-petrov-d}.
Next we use Lemma \ref{lem:decompose-nabla-w} 
and (\ref{eq:cf-data5}) to find the orthogonal splitting of $\lambda_a$. This enables us to compute 
the orthogonal splitting of $Z$, thus proving (\ref{eq:Z-split}). The conclusion of this reasoning is
that the conditions of 
Theorem \ref{theo:kerr-conformal} hold on the initial data hypersurface $\mathcal{S}$
which in turn implies that the conditions of Theorem \ref{theo:kerr-local} hold on 
$\tilde{\mathcal{S}}$. But now we can follow a procedure similar to the proof of Theorem \ref{theo:conformal-petrov-d}
to show that these conditions actually hold in an open set of the data development of $\mathcal{C}$ which contains $\tilde{\mathcal{S}}$.
This is so because after showing that (\ref{eq:type-D-development}) is true one can enlarge the system 
(\ref{eq:propagation-system}) with the following set of equations
\begin{equation}
\pounds_{\vec{\tilde{{\boldsymbol\xi}}}} \mathcal{A}=0\;,\quad  
 \mathcal{A}|_{\tilde{\mathcal{S}}}=0\;,\quad \mathcal{B}|_{\tilde{\mathcal{S}}}<0\;,
\end{equation}
where we use the symbols $\mathcal{A}$, $\mathcal{B}$ to denote any of the quantities
\begin{equation}
\mathcal{A} = \tilde{\Xi}_{a[b}\tilde{\Xi}^*_{c]d}\;,\quad
\mathcal{A} = \mbox{Im}(\tilde{Z}^{3}(\tilde{w}^*)^8)\;,\quad
\mathcal{B} = \frac{\mbox{\em Re}(\tilde{Z}^{3}(\tilde{w}^*)^8)}{\big(18 \mbox{\em Re}\big(\tilde{w}^3\tilde{Z}^*\big)-|\tilde{Z}|^2\big)^3}.
\end{equation}
Under our conditions one can conclude that $\mathcal{A}=0$, $\mathcal{B}<0$ on an open set $\mathcal{U}$ containing $\tilde{\mathcal{S}}$ and so 
Theorem \ref{theo:kerr-local} holds on that open set. Thus $\tilde{g}_{ab}$ is (locally) the Kerr spacetime on
$\mathcal{U}$ and hence the solution $g_{ab}$ of the data $\mathcal{C}$ is conformally related to the Kerr solution.
\qed

\bigskip
\noindent
Theorem \ref{theo:kerr-conformal-data} admits the following converse.
\begin{theorem}
Under the conditions of Theorem \ref{theo:conformal-petrov-d-necessary}
the initial data of any solution of the conformal equations (\ref{eq:cfe0})-(\ref{eq:cfe4})
conformal to the Kerr solution, must comply with (\ref{eq:kerr-conformal-initial-data}) and
conditions (\ref{eq:nut-unphysical-zero})-(\ref{eq:positive-mass-unphysical-2}) 
of Theorem \ref{theo:kerr-conformal} with $Z$ defined by (\ref{eq:Z-split}).
\label{theo:kerr-conformal-data-necessary}
\end{theorem}
\proof 
Since by assumption the unphysical space-time $(\mathcal{M},g_{ab})$ is conformal to the
Kerr solution, then
to show that (\ref{eq:kerr-conformal-initial-data}) and 
(\ref{eq:nut-unphysical-zero})-(\ref{eq:positive-mass-unphysical-2}) hold 
one only needs to find the orthogonal splitting of 
(\ref{eq:killing-property-cfkerr})-(\ref{eq:positive-mass-unphysical-2}) and pull-back the resulting 
conditions to the initial data hypersurface $\mathcal{S}$.
\qed 

\section{The conformal boundary limit}
\label{sec:conformal-limit}
 \begin{figure}[t]
 \centering
 \includegraphics[width=\textwidth,keepaspectratio=true]{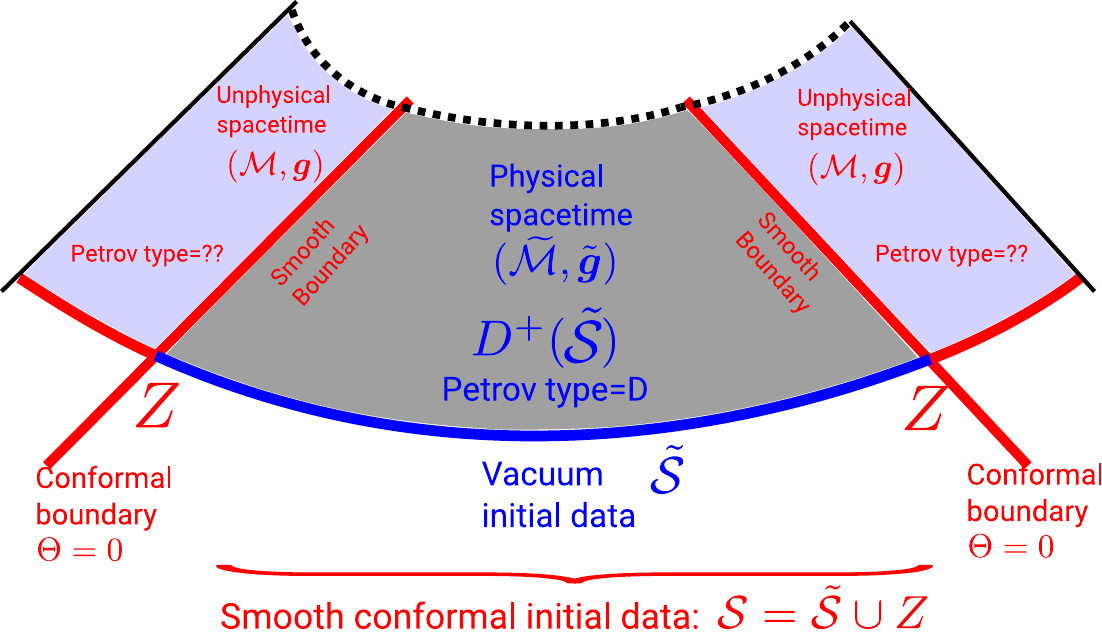}
 % MainResult.pdf: 0x0 px, 300dpi, 0.00x0.00 cm, bb=
 \caption{Data development of the conformal data fulfilling the conditions of Theorem \ref{theo:conformal-petrov-d}.
 \label{fig:conformal-data-dev}}
\end{figure}

Let $\mathcal{C}$ be a data set for the vacuum conformal equations fulfilling the hypotheses of Theorem 
\ref{theo:conformal-petrov-d}. We know then that the solution of these data contains a subset that is 
conformal to a physical type D vacuum solution of the Einstein equations.
However, it is unclear at this point if the unphysical solution will be also of type D outside of 
this subset or its Petrov type will change somewhere (see figure \ref{fig:conformal-data-dev}).
Recall that the Petrov type of a general space-time may change from point to point (see Theorem 7.15
of \cite{GRAHAMHALL}).
In fact the hypotheses of Theorem 
\ref{theo:conformal-petrov-d} on the data $\mathcal{C}$ will be fulfilled
in the region of $\mathcal{S}$ that is mapped under conformal rescaling to the physical space-time data defined on $\tilde{\mathcal{S}}$ 
but it is not clear whether conformal data fulfilling the hypotheses will exist  
outside that region. Therefore the natural question about the existence of a conformal initial data set $\mathcal{C}$ 
meeting the conditions of Theorem \ref{theo:conformal-petrov-d}
outside the region
$\tilde{\mathcal{S}}$ arises. In this section we take a hyperboloidal data set $\mathcal{C}$ fulfilling the hypotheses
of Theorem \ref{theo:conformal-petrov-d} and take the conformal boundary limit $\Omega\rightarrow 0$ of the algebraic and 
differential conditions comprised by Theorem \ref{theo:conformal-petrov-d}. The result (Theorem \ref{theo:data-at-boundary}) 
is that the limit results
in a {\em regular} hyperboloidal conformal initial data set 
at the conformal boundary. That is to say, there are no obstructions to the 
regular extension of the conditions of Theorem 
\ref{theo:conformal-petrov-d} outside $\tilde{\mathcal{S}}$ when the data are hyperboloidal. In particular, 
this implies that the rescaled Weyl tensor $d_{abcd}$ is also of Petrov type D at the conformal boundary 
for hyperboloidal data and therefore the Petrov type is extended at least to the conformal boundary.  
\begin{theorem}[Conformal boundary limit]
Let $\mathcal{C}$ be an initial data set for the vacuum conformal equations fulfilling the 
conditions of Theorem \ref{theo:conformal-petrov-d}, 
and assume further that $h_{ab}$, $K_{ab}$, $\sigma$, $\sigma_a$, and the 
quantities defined from them by \eqref{eq:fund-quantity-1}-\eqref{eq:fund-quantity-6} are all smooth at 
the conformal boundary $\Omega=0$. 
If the data are hyperboloidal then the data set $\mathcal{C}$  
fulfills all the conditions of Theorem \ref{theo:conformal-petrov-d} at $\Omega=0$.
\label{theo:data-at-boundary}
\end{theorem}
\proof
At the conformal boundary $\Omega=0$, the differential 
conditions \eqref{eq:kid-rescaled-1}-\eqref{eq:kid-rescaled-2}
reduce to
\begin{equation}
 \sigma Y h_{ab}=0\;,\quad 
 2\sigma Y_{(b}D_{a)}\Omega+\sigma h_{ab}Y^cD_c\Omega=0.
\label{eq:kid-conformal-boundary}
\end{equation}

Since the data $\mathcal{C}$ are 
by assumption hyperboloidal and regular at the conformal boundary then the metric $h_{ab}$
is non-degenerate at the conformal boundary, $D_c\Omega\neq0$ and $\sigma\neq 0$.
A straightforward computation 
shows that \eqref{eq:kid-conformal-boundary} 
reduces to $Y_a=0$ and $Y=0$ at $\Omega=0$.
Using this information, \eqref{eq:Killing-lapse},
\eqref{eq:Killinglapseshift}
reduce to the following respective conditions at $\Omega=0$.
\begin{eqnarray}
&&2 (\sigma^{b} \sigma^{d} \mathcal{E}_{bd}) + (\sigma_{b} \sigma^{b}) \
\mathit{w} = 0\;,\label{eq:hyp-data-1}\\
&& 
	\lim_{\Omega\rightarrow 0}\frac{Q_{ab}}{\Omega^4}=\nonumber\\
&& 
	2 (\sigma\cdot\sigma+\sigma^2)\mathcal{E}_{ab} 
	- 4 \sigma_{(b}\tilde{\sigma}_{a)} 
	+ 4{\rm i}\; \sigma \mathcal{E}_{(a}{}^{d}\varepsilon_{b)cd} \sigma^{c}
	+ \mathit{w}\sigma_{a} \sigma_{b}\nonumber\\ 
&&
	+ \Bigl(2\sigma\cdot\tilde{\sigma} 
		+ \mathit{w}\bigl(\sigma^2- \sigma\cdot \sigma\bigr) 
	 \Bigr)h_{ab}=0,
\label{eq:hyp-data-3}
\end{eqnarray}
where we introduced the quantities
\begin{equation}
	\tilde{\sigma}_{a}\equiv\sigma^{b} \mathcal{E}_{ab}\;,\quad
	(\sigma\cdot \tilde{\sigma})\equiv \sigma_a\tilde{\sigma}^a\;,\quad
	(\sigma\cdot \sigma) \equiv \sigma_a\sigma^a\;,\quad
	(\tilde{\sigma}\cdot\tilde{\sigma}) \equiv \tilde{\sigma}_a\tilde{\sigma}^a.
\end{equation}

Combining \eqref{eq:hyp-data-1} and \eqref{eq:hyp-data-3} yields 
\begin{equation}
\sigma_a\sigma^a=\sigma^2\;,
\label{eq:hyperboloidal-condition}
\end{equation}
so we have full consistency with the hyperboloidal property of the data. Using this 
condition in the contraction of 
\eqref{eq:hyp-data-3} with $\sigma^b$ we get after using again \eqref{eq:hyp-data-1}
$$
	(2\tilde{\sigma}_{a}  + \mathit{w} \sigma_{a})\sigma+
	2\rm{i}\varepsilon_{acd}\tilde{\sigma}^{d}\sigma^{c} = 0.
$$
The previous equation implies that 
\begin{equation}
2\tilde{\sigma}_{a}  + \mathit{w} \sigma_{a} = 0.
\label{eq:epsilonsigmatransversal}
\end{equation}

Using this condition, \eqref{eq:hyp-data-3} becomes
\begin{equation}
4 \sigma^2 \mathcal{E}_{ab} + 4\mathrm{i} \sigma\sigma^{c} \mathcal{E}_{(a}{}^{d} \varepsilon_{b)cd}
+ 3 \mathit{w}\sigma_{a} \sigma_{b} -  \sigma^2\mathit{w}h_{ab} = 0,
\end{equation}
which can be shown to be equivalent to
\begin{equation}
 4 (\sigma^2 -  \sigma\cdot\sigma) \mathcal{E}_{ab} + 4 \tilde{\sigma}_{(b}\sigma_{a)}
 +\mathit{w} (\sigma\cdot\sigma - \sigma^2) h_{ab} + 2 \mathit{w}\sigma_{b} \sigma_{a} = 0.
\end{equation}
This is trivially fulfilled if the hyperboloidal condition \eqref{eq:hyperboloidal-condition} 
and \eqref{eq:epsilonsigmatransversal} hold. 
It only remains to show that 
the metric $h_{ab}$ is regular at $\Omega=0$. To that end we use again the condition $\mathfrak{a}_{ab}=0$
to obtain the value of the metric. In this case this is 
\begin{equation}
 h_{ab}=2\mathcal{E}_{a}{}^{c}\mathcal{E}_{cb}-\mathcal{E}_{ab}.
\label{eq:h-canonical}
\end{equation}
We recall next, (see e.g. appendix A of \cite{SE-DYNAMICALAWS}) that in an appropriate orthonormal frame, 
the tensor $\mathcal{E}_{ab}$ adopts
the form
\begin{equation}
\mathcal{E}_{ab}=\mbox{diag}(-2z,z,z)\;,\quad z\in\mathbb{C}\;, 
\label{eq:e-canonical}
\end{equation}
and given \eqref{eq:epsilonsigmatransversal} the only possibilities are either $z=w$ or $z=-w/2$. Now using 
these both possibilities in the condition $\mathfrak{a}_{ab}=0$ enables us to find the 
possible values of the metric $h_{ab}$ at the conformal boundary.
Using \eqref{eq:e-canonical} in \eqref{eq:h-canonical} we get
\begin{equation}
 h_{ab}=\frac{24}{\mathit{a}}\mbox{diag}(8z^2+2wz,2z^2-wz,2z^2-wz)\;,
\end{equation}
where according to \eqref{eq:epsilonsigmatransversal}, either $z=-w/2$ or $z=w/4$. Doing the replacements
we find that only the former value gives an orthonormal Riemannian metric. Therefore the metric $h_{ab}$ is regular at the conformal 
boundary if $\sigma_a$ is a simple eigenvalue of $\mathcal{E}_{ab}$.
\qed

The proof of Theorem \ref{theo:data-at-boundary} indicates that, at least for regular hyperboloidal data, there is 
no obstruction to the extension of the Petrov type D from the physical space-time 
to the conformal boundary. 

\section{Conclusions}
\label{sec:conclusions}
We have given necessary and sufficient conditions that a conformal  initial data set 
for the conformal vacuum equations have to satisfy in order that its data development have 
a subset conformal to a type D vacuum solution of the Einstein equations with cosmological constant 
(see Theorems \ref{theo:conformal-petrov-d} and \ref{theo:conformal-petrov-d-necessary} for the complete details). 
In addition we have been able to particularize the results for the case in which the solution of the conformal 
equations is conformal to a suitable region of the Kerr black hole (Theorems \ref{theo:kerr-conformal-data} and 
\ref{theo:kerr-conformal-data-necessary}). 
The conformal data are defined from a vacuum initial data set of the Einstein equations 
with no additional restrictions. This means 
that the data for the conformal equations are constructed in a spacelike hypersurface 
$\tilde{\mathcal{S}}$ which is in the interior of the physical space-time. These data are  
extended to data on a hypersurface $\mathcal{S}\supset\tilde{\mathcal{S}}$, which intersect 
the conformal boundary, by using variables all defined intrinsically on $\mathcal{S}$. 
The regularity of the data so constructed at the conformal boundary
has been also addressed in Theorem \ref{theo:data-at-boundary}. There
we show that there are no obstructions to the extension of the data 
to the conformal boundary when the data are hyperboloidal. 
In particular this implies that the Petrov type can be also extended to
the conformal boundary (in this case it is the Petrov type of the 
rescaled Weyl tensor $d_{abcd}$). 
If the Petrov type D is kept at the conformal boundary then it might be an 
indication that the solution in  
the physical space-time is stable under perturbations.
Recall that hyperboloidal initial data are tied to a conformal
boundary that is null, so there are no obstructions to the extension of 
the Petrov type to the conformal boundary  if it
is null. 

In \cite{PAETZCKID16} (Theorem 3.3) necessary and sufficient 
conditions were found for general data of the vacuum conformal 
equations that 
guarantee that the physical spacetime admits a Killing vector
field . In principle there should be a correspondence 
between the result found in \cite{PAETZCKID16} and 
eqs. \eqref{eq:kid-rescaled-1}-\eqref{eq:kid-rescaled-2}
of our Lemma \ref{lemma:kid-rescaling}. How this correspondence
is actually established is an interesting open question.  

An important aspect that requires further analysis 
is the investigation of existence results for data 
sets fulfilling both the algebraic and the differential 
conditions appearing in Theorems 
\ref{theo:conformal-petrov-d} and 
\ref{theo:conformal-petrov-d-necessary}
and their specializations to the Kerr solution, 
given by Theorems \ref{theo:kerr-conformal-data} and \ref{theo:kerr-conformal-data-necessary}. We 
have
proven in Theorem \ref{theo:data-at-boundary} that for hyperboloidal initial data there are 
no {\em algebraic obstructions} to the extension of the 
conformal initial data described in Theorem 
\ref{theo:conformal-petrov-d} to the conformal boundary but one still
needs to show the existence of actual data fulfilling the conditions
of Theorem \ref{theo:conformal-petrov-d-necessary}. At this point we recall
the work of \cite{FRIEDRICHCF2003,CONFORMALGEODESICSKOTTLER} where 
the existence of congruences of {\em conformal geodesics} 
is proven in globally hyperbolic domains of 
vacuum type D solutions
with a null or timelike conformal boundary. Also interesting in this regard is the work in 
\cite{DIEGOJUANADS2018} where 
conformal initial data for vacuum solutions with a timelike conformal boundary are studied.
All these results imply the existence of a 
{\em conformal extension} giving a conformal boundary with the appropriate properties, 
thus pointing that the conditions of Theorems \ref{theo:conformal-petrov-d} and \ref{theo:conformal-petrov-d-necessary} are going to provide existence results 
for conformal data in a wide range of situations.
For example, an existence result in the case of the Kerr solution of this kind of data would provide 
an important insight into the open problem of the {\em non-linear stability} of this solution. 
It is to be noted that in this paper we have used the standard 1+3 decomposition to formulate the initial value problem
for Friedrich conformal equations, following the approach presented in \cite{DIEGOJUANADS2018,JUANDIEGO2018}. An alternative approach would 
be to use the {\em tractor calculus} in embedded hypersurfaces (see lecture 6 of \cite{CURRYGOVER}). In order that this approach be useful
one would need to find a formulation of Theorem \ref{theo:conf-constraints} in the tractor calculus language.
 Another interesting open question is to find out 
whether the existence results of \cite{ANDERSSON1994,JANOS96} can be adapted in some way to the 
present situation.
The exact extent of all these assertions will be addressed elsewhere.

\section*{Acknowledgements}
We thank Dr. Valiente Kroon for his assistance with the computations 
dealing with the initial value 
formulation of the vacuum conformal equations, 
for a careful reading of the manuscript and for many comments 
and suggestions that improved it.
We thank the financial support from Grant 14-37086G and the
    consecutive Grant 19-01850S of the Czech Science Foundation.
Partial support from the projects
IT956-16 (``Eusko Jaur\-la\-ri\-tza'', Spain),
FIS2014-57956-P (``Ministerio de Econom\'{\i}a y Competitividad'', Spain),
PTDC/MAT-ANA /1275/2014 (``Funda\c{c}\~{a}o para a Ci\^{e}ncia e a Tecnologia'', Portugal)
and the Mobility Fund of the Charles University
is also gratefully acknowledged.

% \bibliographystyle{amsplain}
% \bibliography{/home/alfonso/trabajos/BibDataBase/Bibliography}

\providecommand{\bysame}{\leavevmode\hbox to3em{\hrulefill}\thinspace}
\providecommand{\MR}{\relax\ifhmode\unskip\space\fi MR }
\providecommand{\MRhref}[2]{%
  \href{http://www.ams.org/mathscinet-getitem?mr=#1}{#2}
}
\providecommand{\href}[2]{#2}

\end{document}